# Site-specific location calibration and validation of ray-tracing simulator NYURay at upper mid-band frequencies

Check for updates

**Mingjun Ying ✉, Dipankar Shakya, Peijie Ma, Guanyue Qian & Theodore S. Rappaport ✉**

Ray-tracing (RT) simulators are essential for wireless digital twins, enabling accurate site-specific radio channel prediction for next-generation wireless systems. Yet, RT simulation accuracy is often limited by insufficient measurement data and a lack of systematic validation. This paper presents site-specific location calibration and validation of NYURay, NYU's in-house ray tracer, at upper mid-band frequencies (6.75 GHz and 16.95 GHz). We propose a location calibration algorithm that corrects GPS-induced position errors by optimizing transmitter-receiver (T-R) locations to align simulated and measured power delay profiles, improving T-R location accuracy by 42.3% for line-of-sight (LOS) and 13.5% for non-line-of-sight (NLOS) scenarios. Validation across 18 T-R locations over distances from 40 to 880 m between transmitter and receiver shows excellent RT accuracy in path loss prediction, with path loss exponent (PLE) deviations under 0.14 when comparing measured versus RT predicted power at all T-R locations. While RT underestimates measured delay spread and angular spreads, their cumulative distributions remain statistically similar. The validated NYURay advances RT validation and provides reliable channel statistics for 6G deployment.

Ray tracing (RT) simulation provides deterministic wireless channel prediction for diverse applications spanning network planning, coverage optimization, interference analysis, beamforming design, position location, and channel modeling for emerging technologies, including reconfigurable intelligent surfaces and integrated sensing and communications[1–4]. Site-specific propagation channel prediction enables operators and researchers to optimize base station placement, predict coverage gaps, and validate network performance, which typically proves impractical to analyze empirically before deployment[5,6]. Early work in RT demonstrated the feasibility of site-specific prediction for microcellular environments[7] and commercial potential for network design[8], leading to pioneering commercial tools such as SitePlanner and LANPlanner by Wireless Valley Communications that enabled practical site-specific coverage planning and network design[5,6,9].

RT simulation has emerged as the cornerstone technology for deterministic wireless channel prediction, yet fundamental challenges have limited prediction accuracy since initial development in the early 1990s[7,8,10]. The computational complexity of electromagnetic wave propagation modeling, combined with inaccurate material characterization and oversimplified environmental representations, created a persistent gap between theoretical promise and practical deployment[8,10–12]. As wireless networks migrate toward higher frequencies spanning upper mid-band, mmWave, and sub-THz frequencies, RT becomes essential for accurate site-specific channel prediction, as wavelengths approach environmental feature dimensions, making propagation increasingly sensitive to geometric details and material properties[13–15].

Existing RT simulators face critical limitations: (1) computational bottlenecks preventing real-time predictions, (2) inaccurate modeling of propagation mechanisms including diffraction and scattering phenomena, along with insufficient measurement data for RT calibration, (3) inaccurate, inefficient, and insufficiently detailed 3D environmental representations for RT simulation, (4) inadequate material databases lacking frequency-dependent characterization, and (5) poor integration with modern AI/ML frameworks[12,16,17].

Computational bottlenecks have prevented widespread RT deployment despite decades of development efforts. Early work by Seidel and Rappaport[10,11] introduced image theory and ray launching methods but struggled with computational scaling, while early attempts at real-time diffraction modeling focused on 2D parametric formulations[18]. Commercial tools like Wireless InSite and WinProp implemented GPU acceleration yet remained limited by proprietary architectures and lack of comprehensive measurement validation[5]. NVIDIA Sionna achieved significant advances through differentiable ray tracing and GPU acceleration[19], yet lacks integrated measurement data and built-in calibration mechanisms for

NYU WIRELESS, Tandon School of Engineering, New York University, Brooklyn, NY, USA.
✉e-mail: yingmingjun@nyu.edu; tsr@nyu.edu

real-world validation against actual propagation measurements. Recent academic efforts achieved further speedups through novel data structures[20], real-time GPU rendering[21], and CPU/GPU heterogeneous parallelism achieving over 650 × acceleration[22]. However, speed improvements often sacrificed accuracy by omitting higher-order reflections, diffraction, or scattering mechanisms, while critically lacking systematic calibration against measured channel data essential for upper mid-band frequencies and above, motivating our pursuit of a real-time RT solution with comprehensive physics.

Location uncertainty fundamentally undermines RT calibration accuracy, even with advances in material characterization methods. Traditional RT calibration approaches minimize differences between measured and simulated channels through material property adjustment[19,23], yet ignore uncertainty in recorded transmitter and receiver locations. Measurement campaigns typically rely on consumer-grade GPS systems that introduce as much as 5–10 m location errors[24], undermining calibration accuracy and channel model validation. Existing commercial tools and academic frameworks lack integrated functionality for handling location uncertainty[19,25]. To address location uncertainty, this paper introduces a systematic location calibration methodology that aligns measured and RT-simulated omnidirectional power delay profiles through multi-component loss function optimization.

Environmental modeling accuracy for site-specific environments determines RT simulation performance and remains a major constraint. Modeling approaches have evolved from simple geometric primitives to detailed 3D representations, with commercial AutoCAD-based solutions dominating early implementations[5,9]. Modern approaches leverage open-source tools such as Blender and automated generation from databases like OpenStreetMap[1,26], but lack comprehensive material assignment capabilities. The challenge of integrating accurate environmental modeling with comprehensive propagation physics extends beyond geometry to the underlying wave propagation mechanisms, where most RT simulators implement only basic reflection and penetration, neglecting diffraction and scattering mechanisms essential for accurate RT prediction[12].

Material characterization remains inadequately developed for upper mid-band frequencies despite recent advances. Initial RT implementations relied on generic material properties, resulting in errors exceeding 20 dB[16]. Systematic measurement-based characterization established methodologies for implementing frequency-dependent material databases in RT simulations[5,9,27], with recent advances extending calibration techniques to mmWave and sub-THz bands[23]. However, upper mid-band frequencies remain undercharacterized due to limited measurement campaigns[28,29].

High-fidelity measurement datasets are essential for validating RT tools and constructing accurate propagation models. Such datasets enable rigorous benchmarking of simulation accuracy, facilitate calibration of propagation mechanisms, and support the development of robust, generalizable channel models across diverse environments and frequency bands. NYU WIRELESS measurement campaigns provide comprehensive validation datasets across multiple frequency bands essential for RT development. Extensive campaigns at 28 GHz, 73 GHz, and 142 GHz since 2012 enabled development of statistical models, including NYUSIM, and provided calibration data for the original NYURay simulator[23,30–37]. Recent upper mid-band spectrum allocation motivated new measurement campaigns at 6.75 GHz and 16.95 GHz, addressing data scarcity hindering FR1(C) and FR3 channel model development[38–42].

To address the fundamental limitations of computational constraints and environmental characterization, this paper presents an enhanced NYURay simulator that introduces comprehensive architectural improvements and advanced deterministic channel propagation modeling capabilities in contrast to previous NYURay endeavors[23,43]. Advanced propagation mechanisms incorporate ITU-recommended reflection and penetration models, 3D uniform theory of diffraction (UTD)[44], and frequency-dependent rough surface scattering to capture complex propagation phenomena. The modular object-oriented architecture of NYURay facilitates extensibility and customization, while native Blender integration provides a foundation for future automated 3D environment generation and material assignment workflows utilizing the comprehensive 3D creation suite, including modeling tools, material editors, and Python API for automated workflow development[45].

The main contributions are summarized as follows. First, a comprehensive description of the NYURay RT simulator employs four fundamental propagation mechanisms (reflection, penetration, diffraction, and scattering) with GPU-accelerated computation through TensorFlow integration, validated against extensive 6.75 GHz and 16.95 GHz measurements in a detailed 3D environment model of the NYU Brooklyn campus incorporating buildings and urban infrastructure. Second, a multi-step location calibration methodology reduces GPS-induced location uncertainty through alternating minimization of multi-component PDP loss functions, reducing location errors from 5–10 m to sub-meter precision and improving strongest multipath component power prediction in PDP by 0.96 dB for line-of-sight (LOS) and 0.74 dB for non-line-of-sight (NLOS) scenarios on average. Third, validation comparing field measurements at 18 upper mid-band urban outdoor sites to NYURay predictions generated from 3D digital models of each corresponding physical environment demonstrates that NYURay achieves path loss exponent (PLE) differences of only 0.03–0.14 relative to measured values. Following systematic outlier detection using combined ratio and absolute difference criteria, RT and measurement cumulative distributions demonstrate enhanced statistical agreement for root-mean-square (RMS) delay spread and angular spread parameters, though RT simulations underestimate multipath richness compared to empirical observations.

The following sections present comprehensive validation of these contributions through systematic experimental evaluation and analysis, demonstrating NYURay's capabilities for upper mid-band deterministic channel prediction.

## Results

In this section, we demonstrate and validate NYURay, an enhanced Python-based RT simulator, through comprehensive measurement campaigns at upper mid-band frequencies of 6.75 GHz and 16.95 GHz. The validation encompasses location calibration methodology addressing GPS uncertainties, extensive urban microcell (UMi) measurements across 18 transmitter-receiver (T-R) locations with T-R separation distances from 40 to 880 m, and systematic RT performance assessment. Results show excellent path loss prediction accuracy with path loss exponent deviations below 0.14, while revealing characteristic underestimation of delay spread and angular spread parameters. Location calibration achieves 42.28% multipath matching improvement for line-of-sight scenarios and 13.52% for non-line-of-sight conditions. Systematic outlier filtering substantially improves agreement between NYURay simulations and measurements as quantified by Kolmogorov-Smirnov (KS) tests. After outlier filtering, KS statistics decrease and $p$ values increase, yielding statistically indistinguishable distributions for the RMS delay spread and RMS angular spreads (ASA, ASD). Once extreme site T-R pairs are removed, the ray-tracer reliably reproduces the measured spatial dispersion within the expected variability.

### Ray-tracing fundamentals and the need for measurements and calibration

RT emerged as a fundamental technique for wireless channel modeling in the 1980s, with early applications in indoor environments[11] and later expansion to outdoor cellular scenarios[46]. RT provides deterministic channel predictions by modeling EM wave propagation through geometric optics principles, offering significant advantages over empirical models for site-specific analysis[47]. Unlike statistical channel models that rely on repeated random sampling to approximate average channel characteristics, RT captures specific environmental interactions, enabling accurate prediction of channel parameters including path loss (PL), delay spread (DS), and angular spread (AS) for arbitrary T-R configurations in a model of the physical world[10,11].

NYURay, a 3D RT tool developed at NYU WIRELESS, is a simulation platform for wireless propagation across upper mid-band spectrum,

mmWave, and sub-THz bands. The enhanced NYURay simulator builds upon the foundational C++ codebase, originally calibrated using extensive measurements at 28 GHz, 73 GHz, and 142 GHz[23,36,37]. The enhanced simulator has been transitioned to a Python-based platform, expanding beyond basic reflection and refraction to incorporate frequency-dependent propagation mechanisms, including diffraction and scattering, guided by ITU recommendations[44]. A key contribution includes development of an interactive GUI leveraging the NYU standardized point-data format[42] to enable direct comparison between empirical measurements and simulation outputs. The point-data format represents detailed, site-specific information for each T-R location, including key channel statistics (PL, DS, AS). The GUI integrates tools for statistical analysis of channel parameters and provides precise, site-specific validation of RT predictions against measured ground truth.

NYURay simulates wireless propagation by modeling EM wave interactions with the environment. For TX and RX placed in a propagation environment containing obstacles, NYURay traces rays from TX and calculates various paths to RX through reflection, penetration (transmission), diffraction, and scattering mechanisms[17].

For each individual propagation path $p$, NYURay computes the electric field contribution $\mathbf{E}_p(\mathbf{r}_R)$ at RX location $\mathbf{r}_R$ using:

$$\mathbf{E}_p(\mathbf{r}_R) = \frac{\lambda}{4\pi} \cdot \frac{e^{-jkd_p}}{d_p} \cdot \mathbf{F}_R \cdot \mathbf{T}_p \cdot \mathbf{F}_T \cdot \mathbf{E}_0 \tag{1}$$

where $\lambda$ represents wavelength and $k = 2\pi/\lambda$ denotes wavenumber. The total path length equals $d_p$, while $\mathbf{E}_0$ represents the initial complex electric field vector. Matrices $\mathbf{F}_T$ and $\mathbf{F}_R$ characterize TX and RX antenna field patterns in departure and arrival directions. The cumulative transition matrix $\mathbf{T}_p$ encapsulates all electric field transformations along the path. The expression $\frac{e^{-jkd_p}}{d_p}$ accounts for free-space PL and phase shift, while $\frac{\lambda}{4\pi}$ relates far-field radiation to source characteristics.

The total electric field at RX combines contributions from all propagation paths through coherent vector summation, expressed in the Master Propagation Equation:

$$\begin{aligned}\mathbf{E}_{\text{total}}(\mathbf{r}_R) = \underbrace{\mathbf{E}_L(\mathbf{r}_R)}_{\text{LOS}} &+ \underbrace{\sum\nolimits_{k\in\mathcal{R}} \mathbf{E}_{R,k}(\mathbf{r}_R)}_{\text{Reflected Paths}} + \underbrace{\sum\nolimits_{\ell\in\mathcal{T}} \mathbf{E}_{T,\ell}(\mathbf{r}_R)}_{\text{Transmitted Paths}} \\ &+ \underbrace{\sum\nolimits_{m\in\mathcal{D}} \mathbf{E}_{D,m}(\mathbf{r}_R)}_{\text{Diffracted Paths}} + \underbrace{\sum\nolimits_{n\in\mathcal{S}} \mathbf{E}_{S,n}(\mathbf{r}_R)}_{\text{Scattered Paths}}.\end{aligned} \tag{2}$$

where $\mathbf{E}_L(\mathbf{r}_R)$ represents the LOS component (when unobstructed). The remaining terms incorporate paths categorized by dominant propagation mechanisms: reflection ($\mathcal{R}$), transmission/penetration ($\mathcal{T}$), diffraction ($\mathcal{D}$), and scattering ($\mathcal{S}$). Each category contributes field components $\mathbf{E}_{R,k}$, $\mathbf{E}_{T,\ell}$, $\mathbf{E}_{D,m}$, and $\mathbf{E}_{S,n}$, respectively. The coherent vector summation captures interference effects among all paths at RX location.

NYURay implements four fundamental propagation mechanisms: (1) Reflection computed using a hybrid approach combining Shooting and Bouncing Rays[23], image-based RT, and Fresnel reflection coefficients calculated based on material properties, polarization, and incidence angle[44]; (2) Penetration modeled for waves transitioning between media with overall loss, accounting for both boundary reflections and internal absorption[44]; (3) Diffraction implemented using Uniform Theory of Diffraction (UTD)[48] where rays emanate from edges along Keller cones[49] with computed diffraction coefficients dependent on incidence angle, observation angle, and edge geometry; and (4) Scattering applied for rough surfaces exceeding the Rayleigh criterion[13], with scattered energy distribution following the Lambertian pattern[50].

Accurate 3D environment models enable reliable RT simulations and subsequent T-R location calibration by capturing the geometric and electromagnetic characteristics of the propagation environment. For outdoor UMi 3D environment modeling, a systematic four-stage procedure ensures high-fidelity geometric accuracy and electromagnetic material characterization:

**Stage 1—Initial model import:** OpenStreetMap (OSM) geographical data, including terrain elevation and building footprints, is imported using the BLOSM addon for Blender[51]. OSM provides basic building outlines with typical height estimates, establishing ground reference coordinates and initial geometric framework. However, OSM data often contains dimensional inaccuracies and lacks detailed structural information required for precise RT modeling.

**Stage 2—Ground-truth measurement and model refinement:** Building dimensions undergo systematic correction using a dual-measurement approach with high-precision equipment. The Bosch DLR165 Digital Laser Rangefinder (±1/16" or 1.5 mm accuracy) serves as the primary measurement tool for structural dimensions, including building heights, hallway heights, corridor heights, lamp post heights, and other critical vertical/horizontal distances. iPhone 16 Pro LiDAR scanning (256 × 192 pixel resolution, reliable up to 10 m range) provides cross-validation for smaller urban furniture objects, including trash bins, fire hydrants, electrical transformers, benches, and street-level features, with approximately 1 cm accuracy. This dual-measurement methodology combines the laser rangefinder's high-precision absolute measurements with LiDAR's capability to verify complex object geometries and relative positioning. Each building undergoes comprehensive measurement verification, with critical dimensions measured from multiple vantage points to ensure consistency. Manual object creation includes detailed modeling based on these validated physical measurements and photographic documentation (see Fig. 1 for real-world versus modeled object comparisons).

**Stage 3—Material property assignment:** Object receives appropriate material characterizations reflecting frequency-dependent electromagnetic properties. Material properties follow ITU-R P.2040-3, "Discussion", Table 3, with frequency-specific values interpolated for 6.75 GHz and 16.95 GHz.

**Stage 4—Model export and validation:** The finalized 3D model exports to Mitsuba XML format for NYURay RT electromagnetic field computations. Model validation compares key geometric parameters against ground-truth measurements, achieving < 10 cm RMS error for building dimensions and <5 cm accuracy for object placement. The refined environmental model provides the geometric and electromagnetic foundation enabling accurate RT predictions and reliable location calibration results.

Figure 2 shows the isometric view of the detailed UMi 3D model. Precise geometry (building dimensions, object placement) and correct material properties minimize location calibration algorithm requirements to compensate for modeling inaccuracies. Detailed environmental features ensure simulated MPCs match real-world measurements, enabling reliable calibration results.

Environmental 3D models capture geometric details and material properties with precision for accurate EM field calculations. However, discrepancies between simulated and measured channels arise from modeling limitations, computational approximations, and environmental uncertainties. Calibration bridges the gap between theoretical predictions and real-world propagation by systematically adjusting simulation parameters to match empirical observations.

Calibration to high-quality measurement data is essential for RT tools because: (1) theoretical scattering models may not capture all physical mechanisms in real environments; (2) material property databases contain uncertainties affecting reflection and transmission coefficients; (3) geometric approximations in 3D models introduce systematic errors; and (4) computational limitations necessitate accuracy-efficiency trade-offs. Extensive measurement campaigns provide ground truth data for validating and refining RT predictions, ensuring simulation tools deliver reliable channel characterization for wireless system design[10,13,19,23,36,44,50].

## Measurement campaign

The NYU WIRELESS measurement campaign characterized propagation in FR1(C) and FR3 bands at 6.75 GHz and 16.95 GHz. The campaign focused on outdoor UMi, InH, and InF environments[39–42,52–55]. Upper mid-band measurements at 6.75 GHz and 16.95 GHz were meticulously collected and processed at NYU WIRELESS.

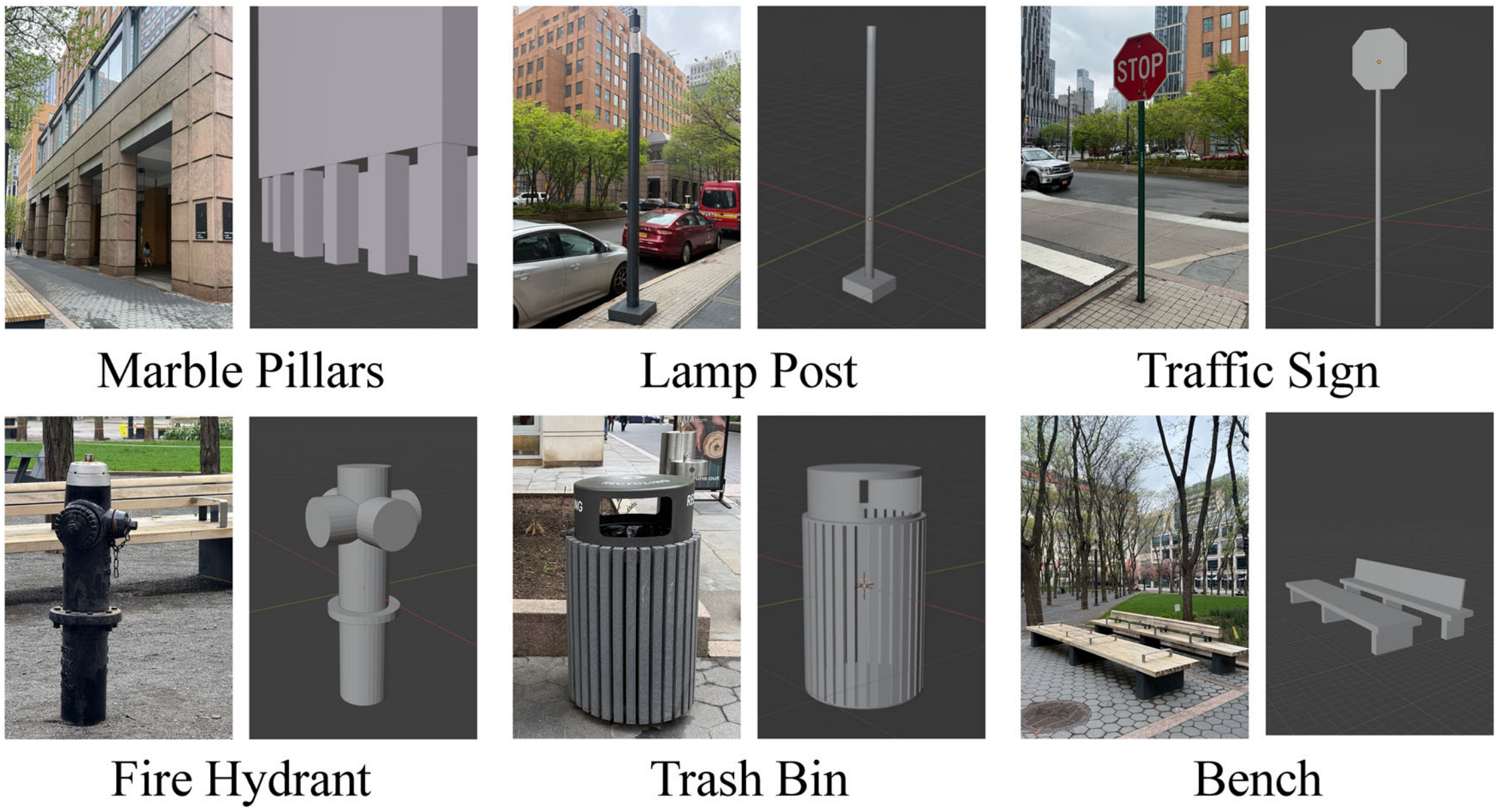

**Fig. 1 | Comparison of real environmental objects and corresponding 3D models in the Downtown Brooklyn, NY UMi measurement area around the NYU Brooklyn campus.** Objects include marble pillars, lamp posts, traffic signs, fire hydrants, trash bins, and benches. Additional urban facilities, such as transformers, trees, and other environmental objects not shown, were included in the comprehensive 3D environmental model for NYURay RT simulation, using precise locations and dimensions from field measurements.

**Measurement system**: The FR1(C) and FR3 measurement campaign utilized a sliding-correlation channel sounder operating at 6.75 GHz and 16.95 GHz, detailed in refs. 40,52. The sounder provides 1 GHz BW, enabling 1 ns temporal resolution. Key components include dual-band co-located front-ends and steerable horn antennas (15 dBi, 30° HPBW at 6.75 GHz; 20 dBi, 15° HPBW at 16.95 GHz). Further system specifications, including PN sequence design and block diagrams, are available in refs. 13,40,56,57. An operational EIRP of 31 dBm was maintained, providing link margins of 142.7 dB at 6.75 GHz and 143.8 dB at 16.95 GHz[39,41]. Antennas supported V/H polarization.

**Measurement data collection procedure:** Multipath data collection involved a two-step process: rapid RX scans to identify significant TX AODs, followed by detailed RX azimuthal sweeps in HPBW steps for comprehensive MPC capture at each T-R location[58]. The procedure, including five RX azimuthal sweep configurations per TX AOD and PTP synchronization using a reference MPC[59], captured data for both V-V and V-H polarizations. Omnidirectional PDPs were synthesized from directional measurements[60]. Post-processing involved applying a power threshold to directional PDPs to identify valid MPCs[54]. Each T-R location pair required approximately 4–5 h for measurement due to the meticulous calibration and antenna pointing at both TX and RX[40].

**Location recording and positioning:** Commercial GPS technology recorded geographical coordinates of TX and RX positions throughout the measurement campaign. Location data was captured using iPhone GPS functionality through Google Earth, providing positioning accuracy of approximately 5–10 m. GPS location uncertainty in the urban canyon of downtown Brooklyn represents a fundamental limitation when correlating site-specific measurements with RT simulations, as small location discrepancies can significantly impact multipath propagation characteristics in dense urban environments[17].

For accurate T-R separation distance determination, laser ranging measurements were employed in addition to GPS location. Laser-based distance measurements provided precise T-R separation values independent of GPS coordinate uncertainty, ensuring accurate path loss analysis and model validation. The dual approach leveraged GPS for general positioning context while relying on laser measurements for critical distance-dependent propagation characterization.

To address GPS location logging errors and improve measurement-simulation correlation accuracy, a calibration algorithm was developed and implemented as detailed in Section T-R Location Calibration. The T-R location calibration algorithm compensates for systematic positioning errors and enhances the reliability of site-specific comparisons between measured and simulated propagation characteristics. Accurate location calibration is also important for utilizing historical measurement data for enhanced RT material parameter calibration and environment modeling. Real-time kinematic positioning (RTK) offers superior accuracy (1–2 cm) compared to standard GPS (5–10 m). However, practical considerations favor the GPS-based approach: NYU WIRELESS has amassed extensive measurement campaigns over multiple frequency bands using standard GPS, producing valuable cross-frequency datasets; the location calibration methodology leverages and maximizes the value of thousands of existing GPS-based T-R location pairs without requiring costly re-measurement campaigns; dense urban environments frequently cause RTK signal blockages and fix losses, forcing fallback to GPS; and RTK deployment entails additional equipment and antenna requirements that increase cost and reduce accessibility for many research groups worldwide. The location calibration methodology thus provides a pragmatic bridge between legacy GPS-based measurements and accurate RT simulation.

**Measurement environment:** An extensive UMi campaign was conducted around the NYU Tandon School of Engineering campus in MetroTech Commons and on a 1 km path along Myrtle Avenue[39–41,52,53]. Location provides representative dense urban microcell setting with varied building geometries and street furniture, suitable for characterizing propagation at upper mid-band frequencies. The environment included potential reflectors, scatterers, and shadowers such as lamp posts, transformers, concrete and marble pillars, trees, foliage, traffic signs, pedestrians, benches, trash bins, and fire hydrants. Reflections were observed primarily from surrounding buildings and metal lampposts[40]. To emulate typical small-cell outdoor deployments, TXs were positioned at 4 m above ground (similar to lamppost heights, representing small-cell BS), and RXs at 1.5 m (mimicking mobile user equipment). Precise antenna height positioning was achieved using the Bosch DLR165 Digital Laser Rangefinder. The measurement procedure ensures accurate vertical positioning. First, a leveler is used to ensure the horn antenna is strictly aligned and perpendicular to the ground. The laser rangefinder then measures from the center of the horn antenna aperture to the ground. Subsequently, the telescopic mast height is adjusted to achieve the exact target heights of 4 m for TX and 1.5 m for RX antennas. The measurement procedure minimizes vertical positioning errors to within millimeter-level accuracy.

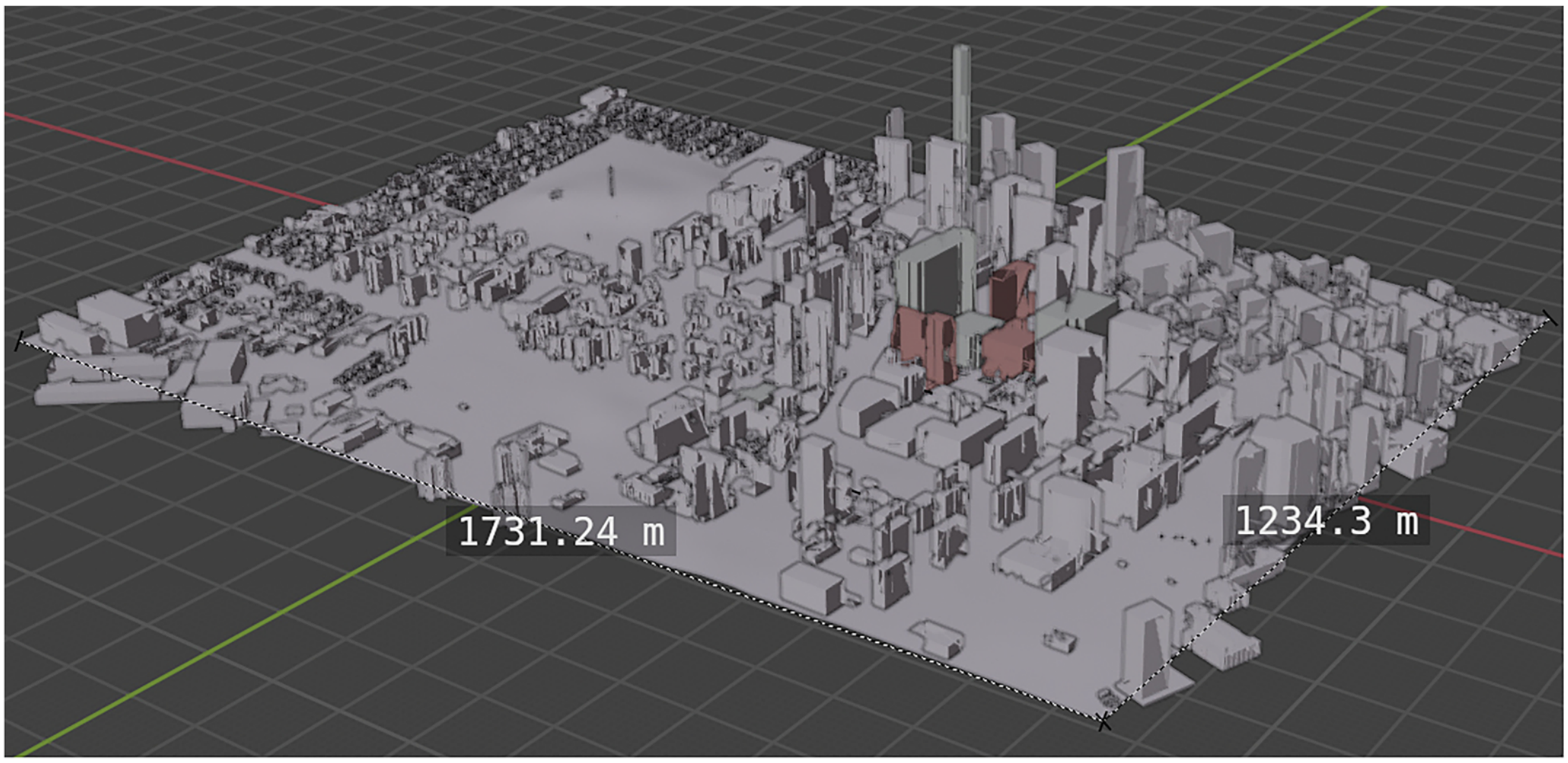

**Fig. 2 | 3D visualization of the Downtown Brooklyn area near the NYU campus used in NYURay simulations, covering approximately 1.2 km × 1.7 km.** Each gray grid represents a 100 m × 100 m area. The model includes buildings, trees, lamp posts, traffic signs, benches, and fire hydrants. The green line pointing toward the bottom-left corner indicates the north direction[17].

TX and RX locations, including T-R pairs in the ~180 m open-square MetroTech Commons and along Myrtle Avenue (a typical street path), are detailed and illustrated in ref. 39. A total of 20 locations were measured (7 LOS, 13 NLOS) with T-R distances from 40 to 1000 m. Two outages at NLOS locations were observed at both 6.75 and 16.95 GHz[39].

**Measurement data processing and point-data format:** Standardized point-data format addresses inconsistent data presentation and facilitates data pooling across organizations[42]. The format represents channel measurements and simulation outputs on a site-specific T-R link basis, retaining full data granularity for accurate environmental depiction at each location pair. Standardizing key channel parameters provides a common framework for direct comparison between empirical measurements and simulation outputs (e.g., from NYURay), enables dataset pooling, and supports statistical analysis or RT validation[42].

Rigorous calibration steps ensured amplitude and delay consistency. Daily field calibrations using a 4 m free-space PL reference corrected fluctuations in output power, oscillator settings, and synchronized Rb clocks[40]. Measurements alternated between V-V and V-H polarizations to mitigate bias and enrich MPC capture. Omni-channel response was synthesized by rotating horn antennas incrementally and combining data in post-processing (removing antenna patterns, summing powers) to emulate an isotropic antenna[60].

The standardized point-data format resolves data format inconsistency across institutions. Standard noise threshold for identifying MPCs from PDPs and spatial threshold for identifying SLs from PAS are consistently implemented[40,55]. The format embeds comprehensive metadata (antenna polarization, site-specific context) alongside raw measurements, providing a common method to display multiple statistics and pool datasets. RMS AS is provided using both 3GPP TR 38.901 identified method[61] and Fleury method[62]. The framework supports pooling for future AI applications and provides machine-readable, interoperable results for validating wireless RTs.

***Statistical Channel Parameter Computation:*** The point-data table values are derived from statistical characterization of key wireless channel parameters, including PL, RMS DS, and RMS AS. Methodologies for computing channel statistics from measured or simulated PDPs follow standard approaches with lognormal distribution assumptions, enabling direct comparison with established 3GPP channel models[61].

*Path Loss Computation:* The CI PL model with 1 m free-space reference distance describes distance-dependent PL using PLE $n$ and shadow fading $\chi_\sigma$[63,64]:

$$\begin{aligned} PL^{CI}(f_c, d_{3D})\ [\mathrm{dB}] &= \mathrm{FSPL}\,(\mathrm{f_c}, 1\mathrm{m}) + 10\mathrm{n}\log_{10}\left(\frac{\mathrm{d_{3D}}}{\mathrm{d_0}}\right) + \chi_\sigma, \\ \mathrm{FSPL}\,(f_c, 1m) &= 32.4 + 20\log_{10}\left(\frac{f_c}{1\ \mathrm{GHz}}\right), \end{aligned} \tag{3}$$

where FSPL($f_c$, 1 m) represents free-space PL at carrier frequency $f_c$ GHz at 1 m, and $\chi_\sigma$ denotes zero-mean Gaussian shadow fading with standard deviation $\sigma^{CI}$ in dB.

*Delay Spread Computation:* RMS DS $\sigma_\tau$ quantifies temporal dispersion as the square root of the second central moment of the PDP[13]:

$$\sigma_\tau = \sqrt{\frac{\sum_{i=1}^{K}(\tau_i - \bar{\tau})^2 P_i}{\sum_{i=1}^{K} P_i}}, \tag{4}$$

where $\bar{\tau} = \frac{\sum_{i=1}^{K}\tau_i P_i}{\sum_{i=1}^{K} P_i}$ represents power-weighted average delay, $\tau_i$ denotes excess delay of the $i$th MPC relative to first arrival, and $P_i$ represents MPC power in linear scale.

*Angular Spread Computation:* RMS AS $\sigma_{AS}$ characterizes spatial dispersion of MPCs. Two primary methods are employed: Fleury's definition and the 3GPP TR 38.901 approach.

Fleury's method[55,62] calculates the second central moment of the PAS:

$$\sigma_{AS}^{Fleury} = \min_{\Delta}\left[\sqrt{\frac{\sum_{n=1}^{N}\sum_{m=1}^{M_n}\left(\theta_{n,m,\mu}(\Delta)\right)^2 \cdot P_{n,m}}{\sum_{n=1}^{N}\sum_{m=1}^{M_n} P_{n,m}}}\right], \tag{5}$$

where $\theta_{n,m,\mu}(\Delta) = \bmod\,(\theta_{\mathrm{n,m}}(\Delta) - \mu(\Delta) + \pi, 2\pi) - \pi$ represents relative direction, $\mu_\theta(\Delta) = \frac{\sum_{n=1}^{N}\sum_{m=1}^{M_n}\theta_{n,m}(\Delta)\cdot P_{n,m}}{\sum_{n=1}^{N}\sum_{m=1}^{M_n} P_{n,m}}$ denotes power-weighted mean angle, and $\Delta$ represents angular shift array.

**Table 1 | Point-data statistics for UMi large-scale spatio-temporal RT simulations, spatially averaged over 25 RX locations (one central point plus eight radial directions at one, two, and three wavelengths), at 6.75 GHz and 16.95 GHz[41,42]**

| Freq. [GHz] | TX | RX | Loc Type | TR Sep [m] | Omni Abs. PL [dB] | Omni DS [ns] | Omni AoA 3GPP (6) [°] | Omni AoA Fleury (5) [°] | Omni AoD 3GPP (6) [°] | Omni AoD Fleury (5) [°] | Omni ZoA 3GPP (6) [°] | Omni ZoA Fleury (5) [°] | Omni ZoD 3GPP (6) [°] | Omni ZoD Fleury (5) [°] |
|---|---|---|---|---|---|---|---|---|---|---|---|---|---|---|
| **6.75** | TX1 | RX1 | LOS | 40 | 77.05 | 13.45 | 14.30 | 15.07 | 15.95 | 18.26 | 6.23 | 6.24 | 3.32 | 3.33 |
| | | RX2 | LOS | 100 | 85.29 | 23.53 | 6.69 | 7.66 | 11.76 | 16.02 | 3.37 | 3.38 | 2.79 | 2.79 |
| | | RX3 | LOS | 193 | 91.29 | 82.08 | 1.19 | 1.64 | 27.60 | 40.30 | 1.91 | 1.91 | 1.65 | 1.65 |
| | | RX4 | NLOS | 561 | 103.55 | 80.72 | 0.46 | 0.48 | 32.47 | 44.71 | 0.42 | 0.42 | 0.88 | 0.88 |
| | | RX5 | NLOS | 880 | 113.59 | 4.06 | 2.44 | 2.45 | 1.33 | 1.98 | 0.81 | 0.81 | 0.44 | 0.44 |
| | | RX7 | LOS | 420 | 99.29 | 12.39 | 0.67 | 0.68 | 13.21 | 18.96 | 0.61 | 0.61 | 2.08 | 2.09 |
| | TX2 | RX1 | LOS | 58 | 82.25 | 14.13 | 17.53 | 17.53 | 15.04 | 18.47 | 2.40 | 2.40 | 3.42 | 3.49 |
| | | RX2 | NLOS | 51 | 84.39 | 55.76 | 25.04 | 29.21 | 18.01 | 19.41 | 4.41 | 4.43 | 3.17 | 3.17 |
| | | RX3 | NLOS | 101 | 102.02 | 22.74 | 19.14 | 19.50 | 2.95 | 2.97 | 2.03 | 2.03 | 0.69 | 0.69 |
| | TX3 | RX1 | NLOS | 71 | 124.76 | 68.33 | 13.69 | 13.75 | 4.90 | 4.89 | 1.75 | 1.75 | 1.94 | 1.94 |
| | | RX2 | NLOS | 45 | 100.45 | 12.13 | 2.33 | 2.33 | 17.19 | 18.81 | 0.91 | 0.91 | 0.47 | 0.47 |
| | | RX3 | NLOS | 125 | 117.72 | 1.24 | 0.69 | 0.43 | 0.35 | 0.17 | 0.28 | 0.17 | 0.51 | 0.30 |
| | TX4 | RX1 | NLOS | 89 | 102.15 | 98.68 | 22.40 | 23.88 | 41.93 | 49.01 | 2.17 | 2.17 | 3.63 | 3.65 |
| | | RX2 | LOS | 120 | 88.91 | 18.07 | 9.34 | 11.60 | 18.43 | 19.76 | 2.74 | 2.75 | 4.37 | 4.38 |
| | | RX4 | NLOS | 184 | 146.98 | 29.24 | 6.27 | 6.27 | 4.18 | 4.18 | 1.03 | 1.03 | 2.06 | 2.06 |
| | TX5 | RX1 | LOS | 52 | 80.53 | 32.78 | 18.55 | 18.91 | 17.80 | 18.29 | 3.90 | 3.90 | 2.39 | 2.39 |
| | | RX2 | NLOS | 167 | 106.99 | 7.93 | 4.41 | 4.42 | 9.61 | 9.70 | 1.10 | 1.10 | 1.71 | 1.71 |
| | | RX3 | NLOS | 141 | 115.64 | 334.65 | 40.97 | 39.50 | 44.61 | 43.08 | 1.83 | 1.83 | 1.82 | 1.82 |
| **16.95** | TX1 | RX1 | LOS | 40 | 85.50 | 10.71 | 12.09 | 12.56 | 13.76 | 15.23 | 6.02 | 6.03 | 2.97 | 2.98 |
| | | RX2 | LOS | 100 | 93.51 | 19.70 | 6.01 | 6.77 | 9.30 | 12.39 | 3.32 | 3.33 | 2.52 | 2.52 |
| | | RX3 | LOS | 193 | 99.56 | 81.79 | 0.87 | 1.13 | 27.02 | 39.60 | 1.94 | 1.94 | 1.38 | 1.38 |
| | | RX4 | NLOS | 561 | 115.53 | 148.37 | 0.54 | 0.56 | 71.04 | 72.83 | 0.46 | 0.46 | 0.93 | 0.93 |
| | | RX5 | NLOS | 880 | 124.40 | 4.00 | 2.16 | 2.17 | 1.19 | 1.85 | 0.77 | 0.77 | 0.39 | 0.39 |
| | | RX7 | LOS | 420 | 108.19 | 8.81 | 0.56 | 0.56 | 11.00 | 15.67 | 0.51 | 0.51 | 1.82 | 1.84 |
| | TX2 | RX1 | LOS | 58 | 90.01 | 13.04 | 19.35 | 19.27 | 11.38 | 13.73 | 2.72 | 2.72 | 2.64 | 2.69 |
| | | RX2 | NLOS | 51 | 94.66 | 69.60 | 26.12 | 29.95 | 18.80 | 20.28 | 4.04 | 4.06 | 3.08 | 3.08 |
| | | RX3 | NLOS | 101 | 110.05 | 21.26 | 18.93 | 19.30 | 2.72 | 2.73 | 2.02 | 2.02 | 0.69 | 0.69 |
| | TX3 | RX1 | NLOS | 71 | 136.67 | 61.19 | 14.75 | 14.75 | 6.06 | 6.06 | 1.65 | 1.65 | 2.03 | 2.03 |
| | | RX2 | NLOS | 45 | 108.47 | 10.14 | 2.17 | 2.18 | 18.49 | 20.27 | 0.97 | 0.97 | 0.46 | 0.46 |
| | | RX3 | NLOS | 125 | 130.89 | 0.28 | 0.32 | 0.10 | 0.12 | 0.04 | 0.10 | 0.04 | 0.18 | 0.07 |
| | TX4 | RX1 | NLOS | 89 | 110.68 | 97.80 | 19.83 | 21.20 | 35.09 | 44.75 | 2.00 | 2.00 | 2.71 | 2.73 |
| | | RX2 | LOS | 120 | 97.30 | 12.08 | 7.54 | 8.85 | 15.86 | 16.90 | 2.55 | 2.56 | 3.76 | 3.77 |
| | | RX4 | NLOS | 184 | 142.97 | 19.89 | 3.46 | 3.45 | 3.02 | 3.05 | 0.55 | 0.55 | 1.11 | 1.10 |
| | TX5 | RX1 | LOS | 52 | 88.68 | 32.27 | 16.79 | 17.24 | 17.59 | 18.02 | 3.85 | 3.85 | 2.20 | 2.20 |
| | | RX2 | NLOS | 167 | 114.58 | 7.93 | 4.23 | 4.24 | 10.06 | 10.24 | 1.17 | 1.17 | 1.73 | 1.73 |
| | | RX3 | NLOS | 141 | 124.80 | 275.08 | 32.01 | 32.33 | 34.61 | 35.35 | 1.61 | 1.61 | 1.46 | 1.47 |

The corresponding point-data table for UMi measurement results can be found in ref. 39.

The 3GPP TR 38.901 method[55,61] employs circular standard deviation using complex phasors:

$$\sigma_{AS}^{3GPP} = \sqrt{-2\times \ln\left|\frac{\sum_{n=1}^{N}\sum_{m=1}^{M_n} e^{(j\theta_{n,m})}\cdot P_{n,m}}{\sum_{n=1}^{N}\sum_{m=1}^{M_n} P_{n,m}}\right|}, \quad (6)$$

where $\theta_{n,m}$ represents angle of mth MPC in nth cluster with power $P_{n,m}$. The method eliminates phase wrapping ambiguities and computational overhead of Fleury's approach.

The point-data tables (e.g., Table 1) enable site-specific analysis and statistical modeling, allowing CDF generation or curve fitting directly from the table for broader 3GPP and NGA applicability[42]. Tables include detailed RT simulation parameters, including center frequency, TX/RX locations, PL (Omni Abs. PL spatially averaged over 25 locations: the original calibrated RX point, plus points shifted outward by one, two, and three wavelengths along eight uniformly spaced directions in the horizontal plane, including the four cardinal directions and four diagonal directions), DS (Omni RMS DS), and AS (Omni RMS ASA/ASD/ZSA/ZSD). RMS AS computation uses both Fleury[62] and 3GPP methods[61]. RMS DS are calculated following the method in 3GPP TR 38.901[61].

### T-R location calibration

The T-R location calibration method optimizes transmitter and receiver positions for each individual T-R link independently by minimizing discrepancies between simulated and measured PDPs. Each T-R pair undergoes separate calibration, where the algorithm refines the TX and RX positions for that specific link without affecting other T-R pairs. This per-link approach ensures site-specific accuracy while maintaining computational tractability. The calibration addresses GPS-induced location uncertainty in field measurements, enhancing RT validation accuracy through systematic location refinement.

**Problem formulation:** For a given T-R pair under calibration, initial TX and RX locations $\mathbf{p}_{TX}^{(0)}$ and $\mathbf{p}_{RX}^{(0)}$, derive from GPS coordinates recorded during measurements. GPS coordinates transform into local Cartesian coordinates aligned with the 3D environment model using Azimuthal Equidistant (AEQD) projection[65]. Initial z-coordinates correspond to antenna heights: 4 m for TX and 1.5 m for RX.

$$\mathbf{p}_{TX}^{(0)} = \left[x_{TX}^{(0)}, y_{TX}^{(0)}, z_{TX}^{(0)}\right], \ \mathbf{p}_{RX}^{(0)} = \left[x_{RX}^{(0)}, y_{RX}^{(0)}, z_{RX}^{(0)}\right]. \tag{7}$$

Location optimization determines TX and RX position adjustments $\Delta\mathbf{p}_{TX}$ and $\Delta\mathbf{p}_{RX}$:

$$\mathbf{p}_{TX}^{*} = \mathbf{p}_{TX}^{(0)} + \Delta\mathbf{p}_{TX}, \ \mathbf{p}_{RX}^{*} = \mathbf{p}_{RX}^{(0)} + \Delta\mathbf{p}_{RX}. \tag{8}$$

For each T-R pair, optimal locations minimize PDP discrepancies between the simulated and measured omnidirectional PDPs for that specific link:

$$\{\Delta\mathbf{p}_{TX}^{*}, \Delta\mathbf{p}_{RX}^{*}\} = \arg \min_{\Delta\mathbf{p}_{TX}, \Delta\mathbf{p}_{RX}} L(P_{sim}, P_{meas}), \tag{9}$$

where $L(\cdot, \cdot)$ quantifies differences between simulated omnidirectional PDP $P_{sim}$ and measured omnidirectional PDP $P_{meas}$[39,60].

Physical constraints limit the position adjustment search space:

$$\| \Delta\mathbf{p}_{TX} \| \leq d_{\max}, \ \| \Delta\mathbf{p}_{RX} \| \leq d_{\max}, \ \Delta z_{TX} = \Delta z_{RX} = 0, \tag{10}$$

where $d_{\max} = 10$ m represents maximum T-R location adjustment based on GPS accuracy limitations[24]. Height constraints apply since vertical antenna positions are measured with laser rangefinders. Accurate antenna height measurement is critical for effective calibration results. Height errors would cause fundamental mismatches between measured and simulated multipath components due to misaligned AOA and AOD, making the PDPs compare substantially different multipath scenarios. Height errors would compromise the calibration algorithm ability to find optimal T-R positions. The fixed height constraint of $\Delta z_{TX} = \Delta z_{RX} = 0$ enables efficient 2D optimization, whereas inaccurate heights would necessitate computationally expensive 3D optimization, including height adjustments.

**RT simulation and PDP generation:** NYURay models EM wave propagation, including LOS paths, reflections, scattering, and diffractions. For each T-R pair, NYURay computes paths with complex amplitudes $a_i$ and delays $\tau_i$. The CIR formulation:

$$h(t) = \sum_{i=1}^{N_p} a_i \delta(t - \tau_i), \tag{11}$$

where $N_p$ denotes path count and $\delta(\cdot)$ represents the Dirac delta function. PDP calculation follows:

$$P(\tau) = |h(t)|^2 = \sum\nolimits_{i=1}^{N_p} |a_i|^2 \delta(t - \tau_i). \tag{12}$$

**PDP alignment techniques:** Temporal alignment compensates for reference point differences and timing offsets between simulated and measured PDPs. Maximum peak alignment utilizes dominant path delays:

$$\Delta\tau_{simple} = \tau_{sim}^{\max} - \tau_{meas}^{\max}, \tag{13}$$

where $\tau_{sim}^{\max}$ and $\tau_{meas}^{\max}$ denote delays of maximum power in simulated and measured PDPs. For complex multipath environments, correlation-based alignment applies:

$$\Delta\tau_{multi} = \arg \max_{\Delta\tau} \ \mathrm{corr}\,(P_{sim}(\tau), P_{meas}(\tau + \Delta\tau)). \tag{14}$$

**Multi-component loss function:** The loss function combines specialized components evaluating PDP similarity:

Peak matching aligns significant MPCs by quantifying temporal and power discrepancies, the peak matching loss ensures that the optimizer finds TX-RX locations where strong MPCs in the simulated PDP align with strong MPCs in the measured PDP, both in arrival time and received power. Accurate MPC alignment is essential because the location calibration algorithm relies on matching the physical propagation paths between simulation and measurement. The peak matching loss quantifies temporal and power discrepancies between matched MPC pairs:

$$L_{peak} = \frac{1}{N_{matched}} \sum_{i=1}^{N_{sim}} \min_j \left[ w_i \left( \frac{|\tau_{sim,i} - \tau_{meas,j}|}{T_{norm}} + |P'_{sim,i} - P'_{meas,j}| \right) \right], \tag{15}$$

where $N_{matched}$ denotes matched peaks; $w_i = e^{-\tau_{sim,i}/\tau_{ref}}$ prioritizes early paths with $\tau_{ref}$ = 500 ns; $T_{norm}$ = 100 ns normalizes delay differences; and $P\prime$, $P\prime$ are normalized peak powers. Measured directional PDPs and synthesized omnidirectional PDPs have absolute timing with sub-nanosecond accuracy, with the PTP-based synchronization employed for the channel sounder[59]. Therefore, the measured PDPs provide an accurate true propagation delay reference (not relative) for aligning simulated PDPs in the ray tracer.

Unmatched peaks penalty prevents the optimizer from converging to TX-RX locations where the simulated PDP contains a different number of significant multipath components than the measured PDP. Without the unmatched peaks penalty, the optimizer might find locations that align one or two strong paths accurately but miss additional paths present in the measurement, or predict spurious paths not observed in the measurement. The penalty quantifies the fractional difference in significant peak counts between simulated and measured PDPs:

$$L_{unmatched} = w_{unmatched} \cdot \frac{|N_{sim} - N_{meas}|}{\max(N_{sim}, N_{meas})}, \tag{16}$$

where $N_{sim}$ and $N_{meas}$ represent significant peak counts in simulated and measured PDPs, and $w_{unmatched}$ = 0.5 weights the penalty relative to other loss components. The normalization ensures the penalty ranges from 0 (identical peak counts) to 1 (one PDP has no peaks matching the other).

The shape loss captures the overall envelope agreement between simulated and measured PDPs across all delay bins, complementing the peak matching loss that focuses only on discrete MPC locations. The shape loss penalizes locations where the simulated PDP has the correct peak positions but incorrect power distribution across the delay domain. The shape loss computes the mean squared error between

dB-scale power values:

$$L_{shape} = \frac{1}{|\mathcal{R}|}\sum_{\tau\in\mathcal{R}}(P_{sim,dB}(\tau) - P_{meas,dB}(\tau))^2, \quad (17)$$

where $P_{dB}$ represents dB-scale power values and $\mathcal{R}$ denotes delay bins exceeding the significance threshold. Computing the shape loss in dB scale ensures that weak multipath components contribute meaningfully to the loss, rather than being dominated by the strongest path in linear scale.

The combined loss function with regularization:

$$L = \alpha\cdot(L_{peak} + L_{unmatched}) + (1-\alpha)\cdot L_{shape} + \beta\cdot L_{distance}, \quad (18)$$

where $\alpha = 0.7$ weights peak-based metrics relative to shape metrics, $L_{distance} = (\| \Delta\mathbf{p}_{TX}\|^2 + \| \Delta\mathbf{p}_{RX}\|^2)/(d^2_{\max})$ penalizes large TX and RX position adjustments from initial GPS coordinates, and $\beta = 0.05$. The combined loss function L in (18) integrates three complementary components: peak-based metrics ($L_{peak} + L_{unmatched}$) that ensure accurate multipath alignment, shape-based metrics ($L_{shape}$) that capture overall PDP envelope similarity, and a distance regularization term ($L_{distance}$) that penalizes large position adjustments from initial GPS coordinates to prevent overfitting. The regularization term constrains the optimization to physically plausible solutions within the known GPS uncertainty bounds.

The multi-component loss function in (18) was developed through systematic experimentation with various formulations to identify the optimal combination for T-R location calibration. Initial attempts using only peak matching loss (15) achieved accurate alignment of dominant multipath components but failed to penalize spurious peaks in simulated PDPs that did not correspond to measured multipaths. Conversely, using shape loss (17) alone captured overall PDP envelope similarity but proved insensitive to precise temporal alignment of individual multipath components critical for location calibration. The unmatched peaks penalty (16) was introduced after observing that optimized locations sometimes produced PDPs with significantly different numbers of resolvable multipaths compared to measurements.

The weighting parameter $\alpha$ = 0.7 in the combined loss function was empirically determined through grid search over α ∈ [0.1, 0.9], with $\alpha = 0.7$ yielding the lowest average calibration error across all T-R pairs. Lower α values ($\alpha < 0.5$) overemphasized shape matching at the expense of precise peak alignment, while higher values ($\alpha > 0.8$) neglected overall envelope agreement. The distance regularization term with $\beta = 0.05$ prevents excessive position adjustments that might overfit to measurement noise; without regularization ($\beta$ = 0), calibrated positions occasionally shifted beyond physically plausible GPS error bounds.

**Location optimization algorithm:** For each individual T-R pair, joint optimization over both TX and RX positions requires searching a four dimensional space, which grows as the fourth power of the grid resolution. The calibration processes one T-R link at a time, optimizing the TX and RX positions for that specific link before proceeding to the next T-R pair. Alternating minimization further reduces computational complexity for each link by optimizing one positionwhile holding the other fixed, reducing the search from one four-dimensional problem to two sequential two dimensionalproblems that grow only as the second power of grid resolution. The optimization employs alternating minimization, sequentially refining the TX position (with RX fixed) then adjusting the RX position (with TX fixed) for the T-R pair under calibration, reducing computational complexity from $O(N^4_{grid})$ to $O(N^2_{grid})$. Each position optimization follows a three-step approach:

**Step 1 Coarse grid search:** Systematic evaluation examines candidate locations within [−5, 5] m range using 2.5 m grid spacing. For each candidate position, NYURay computes the RT simulation and evaluates the loss function $L$. The grid search identifies the coarse optimal region containing the loss minimum.

**Step 2 Fine grid refinement:** Around the coarse optimum, fine grid search employs 0.5 m spacing within ±1.5 m range. The refined search localizes the minimum with sub-meter precision while maintaining computational efficiency.

**Step 3 Gradient-free optimization:** Powell's conjugate direction method[66] performs final refinement from the fine grid optimum. The method iteratively searches along conjugate directions without requiring gradient computation. Gradient-based optimization methods cannot be applied to ray-tracing simulations because ray tracing involves discrete geometric decisions: each multipath component either exists or does not exist depending on whether geometric rays intersect environmental surfaces. Small position changes can cause rays to transition between hitting and missing surfaces, creating step discontinuities in the loss function where gradients do not exist. Powell's method requires only function evaluations without gradient information, making the method suitable for optimizing the non-differentiable RT objective. Multiple iterations refine both positions because the first TX optimization uses the initial (erroneous) RX position, producing a TX estimate that partially compensates for RX error. After RX optimization corrects the RX position, the optimal TX location shifts. Subsequent iterations allow TX and RX positions to converge jointly toward the true locations, with position updates decreasing monotonically until falling below the convergence threshold.

Algorithm 1 details the complete optimization procedure for calibrating a single T-R link. The alternating minimization continues until both TX and RX position updates converge below the threshold $\epsilon = 0.1$ m, typically within 3–5 iterations.

**Algorithm 1**. **T-R Location calibration via alternating minimization for a single T-R pair**

**Require**: Initial positions $\mathbf{p}^{(0)}_{TX}, \mathbf{p}^{(0)}_{RX}$, measured PDP $P_{meas}$, 3D environment model, NYURay RT engine parameters
**Ensure**: Calibrated positions $\mathbf{p}^*_{TX}, \mathbf{p}^*_{RX}$
**Initialize:** $\mathbf{p}_{TX} \leftarrow \mathbf{p}^{(0)}_{TX}, \mathbf{p}_{RX} \leftarrow \mathbf{p}^{(0)}_{RX}$
**Parameters:** $d_{max} = 10$ m, $\epsilon = 0.1$ m, $N_{max} = 10$
**for** $iter = 1$ **to** $N_{max}$ **do**
  $\mathbf{p}^{prev}_{RX} \leftarrow \mathbf{p}_{RX}$
  **Step 1:** Coarse grid search
  **for** $\Delta x \in \{-5, -2.5, 0, 2.5, 5\}$ m **do**
    **for** $\Delta y \in \{-5, -2.5, 0, 2.5, 5\}$ m **do**
      $\mathbf{p}^{cand}_{RX} \leftarrow \mathbf{p}^{(0)}_{RX} + [\Delta x, \Delta y, 0]^T$
      **if** $\| \mathbf{p}^{cand}_{RX} - \mathbf{p}^{(0)}_{RX} \| \le d_{max}$ **then**
        Run RT: $P_{sim} \leftarrow$ NYURay $(\mathbf{p}_{TX}, \mathbf{p}^{cand}_{RX})$
        Evaluate: $L \leftarrow L(P_{sim}, P_{meas})$ using
      **end if**
    **end for**
  **end for**
  $\mathbf{p}^{coarse}_{RX} \leftarrow \arg\min L$
  **Step 2:** Fine grid refinement around $\mathbf{p}^{coarse}_{RX}$
  Search ± 1.5 m with 0.5 m spacing $\rightarrow \mathbf{p}^{fine}_{RX}$
  **Step 3:** Powell optimization from $\mathbf{p}^{fine}_{RX}$
  $\mathbf{p}_{RX} \leftarrow$ Powell $(\mathrm{L}, \mathbf{p}^{\mathrm{fine}}_{\mathrm{RX}}, \mathrm{tol} = 0.1)$
  $\mathbf{p}^{prev}_{TX} \leftarrow \mathbf{p}_{TX}$
  Repeat Steps 1–3 for TX optimization $\rightarrow \mathbf{p}_{TX}$
  **if** $\| \mathbf{p}_{TX} - \mathbf{p}^{prev}_{TX} \| < \epsilon$ **and** $\| \mathbf{p}_{RX} - \mathbf{p}^{prev}_{RX} \| < \epsilon$ **then**
    **break**
  **end if**
**end for**
**return** $\mathbf{p}^*_{TX} = \mathbf{p}_{TX}, \mathbf{p}^*_{RX} = \mathbf{p}_{RX}$

**Grid spacing sensitivity analysis:** The chosen grid spacing configuration (coarse 2.5 m, fine 0.5 m) represents a balanced approach optimizing computational efficiency while maintaining calibration accuracy. The coarse grid contains 25 points (5 × 5) while the fine grid contains 49 points (7 × 7).

Alternative grid configurations demonstrate clear computational trade-offs. Dense grid spacing (0.3 m fine) significantly increases computational time by evaluating 121 points (11 × 11), requiring

## Table 2 | Calibration performance metrics summary

| Scen. | [a]Loss Reduct. (%) | Peak power diff. (dB) Original | Calibrated | [b]Peak power Impr. (dB) |
|---|---|---|---|---|
| LOS | 42.28 ± 9.13 | 2.44 | 2.12 | 0.32 ± 0.09 |
| NLOS | 13.52 ± 5.22 | 8.09 | 7.86 | 0.23 ± 0.11 |

[a]Loss Reduct: Normalized reduction in loss function value (%) using (18).
[b]Peak Power Impr.: Improvement (reduction) in absolute difference between simulated and measured peak power, comparing original RT simulation results to calibrated location results.

approximately 2.5× longer processing time for the grid search phase. Conversely, loose grid spacing (3.0 m coarse) reduces computational time by evaluating only 16 points (4 × 4), cutting grid search time by approximately 40%. However, these time differences become negligible in the final calibration accuracy.

The Powell method gradient-free optimization follows the grid search phase to ensure optimal location convergence regardless of initial grid resolution. Critically, the final Powell optimization guarantees that different grid configurations converge to nearly identical calibrated positions, with peak power predictions differing by less than 0.1 dB across all tested grid resolutions. For example, starting from loose (3.0 m), standard (2.5 m/0.5 m), or dense (0.3 m) grids, the Powell method consistently converges to T-R positions producing peak power predictions within 0.08 dB of each other, demonstrating the robustness of our three-step optimization approach.

The selected 2.5 m/0.5 m configuration balances initial grid search efficiency with subsequent Powell optimization time, minimizing total calibration computational burden while ensuring consistent sub-0.1 dB accuracy in final predictions.

**Calibration performance:** Table 2 presents key performance metrics for the location calibration algorithm 1 that minimizes the combined loss function in (18). Across all measured T-R pairs, calibration yields average reduction of 24.70% in loss function $L$. LOS scenarios demonstrate substantial average loss reduction of 42.28 ± 9.13%, while NLOS scenarios achieve a 13.52 ± 5.22% reduction. For LOS links, average difference between simulated and measured peak power decreases from 2.44 dB to 2.12 dB (0.32 ± 0.09 dB improvement). For NLOS links, average difference decreases from 8.09 dB to 7.86 dB (0.23 ± 0.11 dB improvement).

Figure 3 illustrates the T-R location calibration impact on PDP prediction using NYURay for TX1-RX4 link (560 m separation) at 6.75 GHz. RT simulation using uncalibrated locations overestimates received power by 8 dB and exhibits 4 ns temporal mismatch. RT simulation with calibrated locations reduces delay mismatch to 0.1 ns and power difference to 0.5 dB, demonstrating improved alignment with measured data. The calibration methodology effectively compensates for GPS location errors, enabling accurate site-specific channel validation.

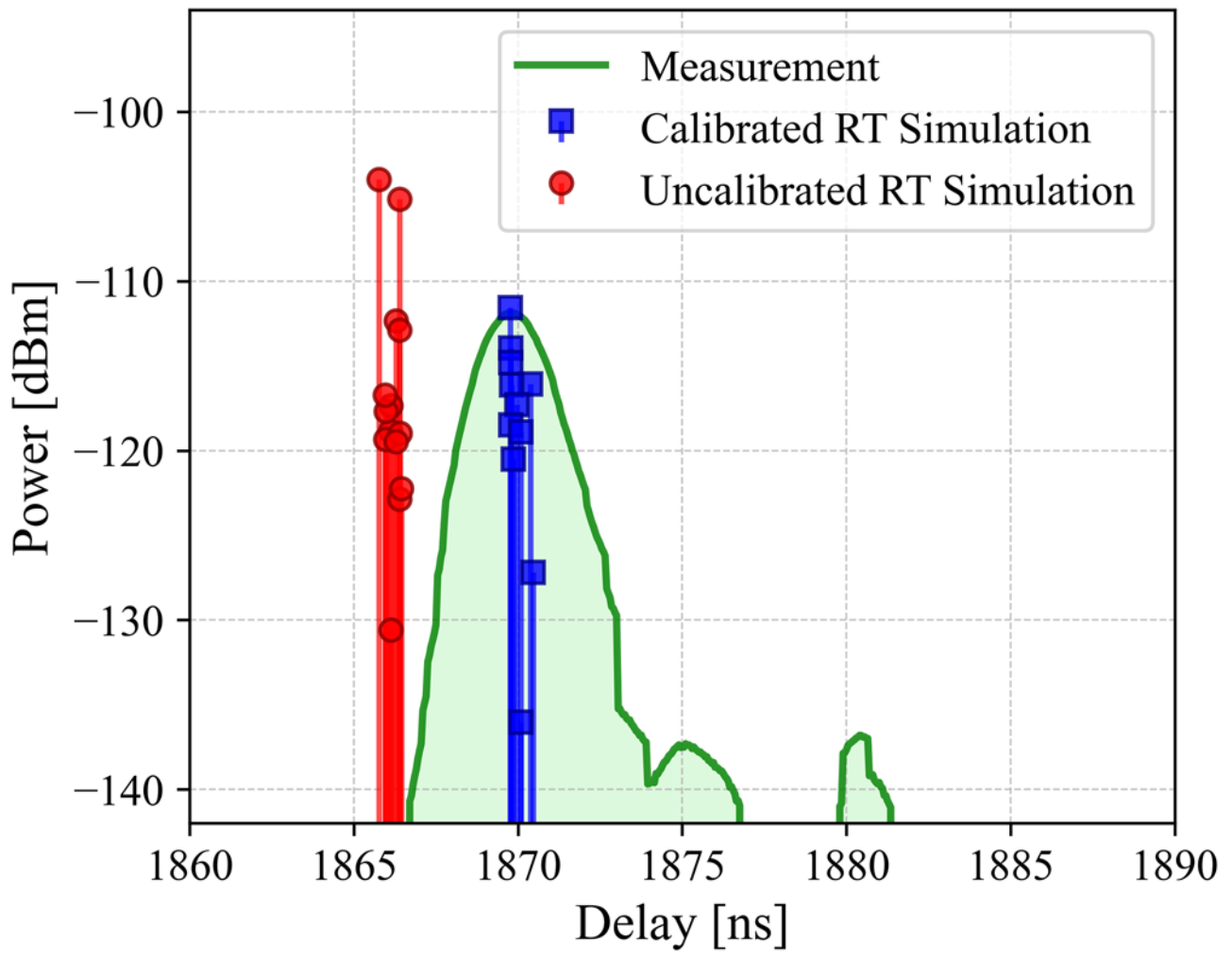

**Fig. 3** | A typical example of PDP comparison showing measured data versus uncalibrated and calibrated location RT simulations for a single T-R link, TX1-RX4 (560 m NLOS), in the Downtown Brooklyn, NY UMi environment at 6.75 GHz[39].

## Table 3 | Key RT simulation parameters for NYURay to evaluate channel prediction accuracy at 6.75 GHz and 16.95 GHz, compared to field measurements conducted in Brooklyn, NY around the NYU Brooklyn campus[39]

| Parameter | Value/Assumption |
|---|---|
| TX/RX antenna pattern | Isotropic, 0 dBi gain |
| Transmit power | 0 dBm |
| Carrier frequency | 6.75 GHz and 16.95 GHz |
| Minimum considered received power | -160 dBm (CIR cutoff threshold) |
| RT mechanisms enabled | Reflection, Diffraction |
| Noise and bandwidth constraints | Not included (idealized CIR used) |

### Ray-tracing simulation setup for validation

To enable direct comparison between RT prediction and field measurement results key RT simulation parameters were configured for NYURay as summarized in Table 3. Isotropic antennas with 0 dBi gain were employed at TX and RX locations for RT prediction. Since field measurements were acquired using directional horn antennas, the measured results were synthesized to emulate omnidirectional antenna characteristics by removing antenna gains, ensuring a fair comparison between measurements and simulations[39,60]. Transmit power was set to 0 dBm for RT simulations at 6.75 GHz and 16.95 GHz. RT simulations used 18 valid T-R locations from the measurement campaign, excluding TX4-RX3 and TX1-RX6 outage locations for both frequencies to maintain consistency between measured and simulated datasets.

A minimum received power threshold (CIR cutoff) of −160 dBm was used in NYURay to match the practical maximum PL of the field measurements of approximately 160 dB[40]. The Mini-Circuits dual-band RF unit provides link margins of 142.7 dB at 6.75 GHz and 143.8 dB at 16.95 GHz with maximum EIRP of 31 dBm. The −160 dBm cutoff ensures weak MPCs remain detectable within the dynamic range. In NYURay, reflection and diffraction were enabled.

Penetration through buildings was disabled due to unavailable detailed building structure information and substantial material penetration losses in outdoor UMi scenarios at studied frequencies[40,53]. Reflection and diffraction dominate outdoor coverage mechanisms.

Scattering effects were neglected based on the Rayleigh criterion[13,67]. According to this criterion, significant diffuse scattering occurs if the RMS surface height variation, $h_{\text{rms}}$, exceeds a threshold height $h_{\text{th}} = \lambda/(8\cos\theta_i)$[67], where $\lambda$ is the wavelength and $\theta_i$ is the angle of incidence relative to the surface normal. The minimum threshold height (the most stringent condition for smoothness) occurs at normal incidence ($\theta_i = 0°$), giving $h_{\text{th,min}} = \lambda/8$. At 6.75 GHz ($\lambda \approx 44.4$ mm), $h_{\text{th,min}} \approx 5.6$ mm. At 16.95 GHz ($\lambda \approx 17.7$ mm) $h_{\text{th,min}} \approx 2.2$ mm.

Typical RMS surface heights for common building materials (Table 4) are in the micrometer range[68]. Comparing the typical RMS heights with the calculated threshold heights (millimeters), we find that $h_{\text{rms}} \ll h_{\text{th}}$ at both 6.75 GHz and 16.95 GHz.

Therefore, the surfaces in the UMi environment are considered electrically smooth at 6.75 GHz and 16.95 GHz, and the Rayleigh criterion for significant diffuse scattering is not met. Consequently, scattering effects were neglected in the RT simulation, simplifying the model while maintaining accuracy for the dominant propagation mechanisms.

NYURay employs ray launching with one million ($1 \times 10^6$) rays per TX location. Ray count balances angular resolution for accurate spatial channel characteristics against computational complexity[10]. Prior RT tests indicate one million rays using hybrid shooting and bouncing ray (SBR) technique with imaging methods provide optimal trade-off for detailed UMi environments[17,23]. Each T-R link simulation using a physical 3D environmental model as shown in Fig. 2 required approximately 2–3 seconds when executed on NVIDIA RTX 4090 GPU (24GB VRAM) with AMD Ryzen Threadripper PRO 7975WX 32-core CPU, providing a reasonable computational time for detailed physics-based modeling, including reflection and diffraction mechanisms. The simulation employs optimized tensor operations in TensorFlow 2.15.1 and efficient ray-geometry intersection algorithms to maintain practical execution times for research applications.

**Table 4 | Typical RMS surface heights for common building materials[68]**

| Material | RMS height $h_{rms}$ (μm) |
|---|---|
| Wood | 76.4–80.0 |
| Plexiglass | 8.2–14.5 |
| Drywall | 96.4–99.2 |
| Concrete | 264.3–269.0 |
| Red brick | 321.3–325.0 |

To contextualize NYURay's performance for research validation, we reference validation studies by Obeidat et al.[69,70] who report Wireless InSite achieves RMSE values of approximately 5 dB at 2.4 GHz and 5.1 dB at 5 GHz for indoor environments, with accuracy heavily dependent on material parameter selection (7–10 dB RMSE variations observed when adjusting conductivity and permittivity values). For computational performance comparison using the NYU Tandon campus environment[39], Sionna v1.0[71] achieves the fastest performance in each T-R location RT simulation with millisecond-level channel predictions through highly optimized GPU parallelization. Wireless InSite[72] offers dual modes: real-time modules enabling millisecond predictions through precomputed ray databases, or full shooting-and-bouncing ray (SBR) requiring 3–5 h per simulation for complex environments. MATLAB Ray Tracing[73] requires approximately 30 s per T-R pair. NYURay achieves path loss RMSE of 3.2 dB for LOS and 5.8 dB for NLOS scenarios at upper mid-band frequencies (6.75 GHz and 16.95 GHz) with simulation times of 2–3 s per T-R pair on NVIDIA RTX 4090 GPU with AMD Ryzen Threadripper PRO 7975WX 32-core CPU. While not the fastest computationally, NYURay provides an optimal balance of accuracy, full physics-based modeling, and reasonable computational time for research applications requiring extensive parametric studies and site-specific channel characterization in upper mid-band propagation and 6G channel modeling.

NYURay calculates power reduction from reflections and diffractions based on interaction geometry and material electromagnetic properties[17]. RT through the 3D environmental model (Fig. 2) accounts for interactions with buildings, ground, and modeled structures using Fibonacci lattice method. Maximum reflection depth was set to 5, allowing up to five reflections per ray sequence while capturing significant MPCs and avoiding excessive computation for highly attenuated higher-order reflections. Received power is calculated by summing contributions of all rays arriving within a receiver sphere using (2)[74].

## Site-specific channel characteristics validation for UMi at upper mid-band

Using the location calibration method in algorithm 1, measurement data collection, and RT simulation settings detailed in previous sections, the validation analysis examines NYURay RT tool performance against measured channel characteristics obtained in the UMi environment at 6.75 GHz and 16.95 GHz[39]. The validation encompasses 18 T-R links comprising 7 LOS and 11 NLOS scenarios, with separation distances ranging from 40 m to 880 m across the NYU Brooklyn campus.

**Comparison of measured and RT simulated path loss:** For each of the 18 simulated T-R links detailed in Table 1, NYURay calculates PL through coherent summation of all multipath components. The complex channel coefficient $h$ represents the coherent sum of all $K$ multipath components:

$$h = \sum_{k=1}^{K} a_k e^{j\phi_k}, \tag{19}$$

where $a_k$ and $\phi_k$ represent the amplitude and phase of the $k$-th MPC. PL in dB follows directly:

$$PL_{\mathrm{dB}} = -10\log_{10}(|h|^2). \tag{20}$$

Table 5 presents comprehensive comparison of CI model parameters extracted from measurements, RT simulations, spatially-averaged RT simulations, and 3GPP standardized models. RT simulations demonstrate accurate PL exponent prediction, particularly for LOS scenarios where differences remain below 0.07 at both frequencies. NLOS predictions show slightly larger deviations (0.10–0.26) but maintain superior accuracy compared to generic 3GPP statistical models that have no site-specific information and merely represent a general environment.

Shadow fading prediction reveals contrasting performance between LOS and NLOS scenarios. LOS measurements exhibit shadow fading standard deviations of 2.57–4.05 dB, while RT simulations produce notably smaller values of 1.01–1.11 dB. The reduced variability stems from deterministic RT modeling capturing dominant propagation paths while missing small-scale environmental variations including pedestrian movement, vehicle presence, and temporal changes in foliage density that contribute to measured shadow fading. NLOS scenarios present an opposite trend, with RT predicting higher shadow fading (11.73–12.01 dB) compared to measurements (6.53–8.78 dB). The overestimation occurs because RT computes signal levels at exact point locations, capturing precise constructive and destructive multipath interference patterns that vary significantly over small spatial distances. In contrast, field measurements inherently perform spatial averaging due to finite antenna aperture and local positional variations during data acquisition, thereby smoothing out extreme signal fluctuations and reducing the observed shadow fading variability.

**Comparison of measured and RT simulated delay spread:** Delay spread validation quantified temporal dispersion characteristics through RMS calculations applied to power delay profiles, employing a −25 dB threshold relative to peak power for significant multipath identification. The analysis compared measured delay spreads with NYURay predictions across 18 T-R locations under both LOS and NLOS conditions.

Table 6 presents validation results revealing systematic patterns in RT delay spread prediction. RT consistently underestimates RMS DS compared to measurements, especially in LOS scenarios, with predicted values of 24.7 ns and 21.6 ns versus measured 62.8 ns and 46.5 ns at 6.75 GHz and 16.95 GHz, respectively. The RT underestimation of RMS DS reflects environmental modeling limitations where the 3D model captures major structures but omits architectural details, window frames, and surface irregularities that create additional scattered paths in real environments. The current NYURay implementation models specular reflections and edge diffraction but excludes diffuse scattering from rough surfaces, contributing to the observed discrepancies.

NLOS DS predictions demonstrate improved agreement with measurements, showing smaller absolute differences (12.2–22.7 ns). The better NLOS performance occurs because dominant propagation mechanisms in obstructed scenarios involve major reflections and diffractions from large structures, which RT effectively captures. The remaining underestimation arises from simplified foliage representation as uniform dielectric blocks, missing complex scattering from branches and leaves, and computational constraints limiting higher-order interaction modeling. Additionally, measurements capture temporal variations from pedestrians, vehicles, and wind-induced foliage movement that create dynamic multipath absent in static RT simulations.

Figure 4 demonstrates the enhanced statistical validation for omnidirectional RMS DS through systematic outlier filtering. For RMS DS,

**Table 5 | Comparison of omnidirectional CI PL exponents ($n$) and shadow-fading standard deviations ($\sigma$) for the UMi environment in Downtown Brooklyn, NY around the NYU Brooklyn campus between 18 T-R locations in NYURay vs. 3GPP**

| Freq. (GHz) | Scenario | Path loss exponent $n$ | | | | | | | Shadowing $\sigma$ (dB) | | | | | | |
|---|---|---|---|---|---|---|---|---|---|---|---|---|---|---|---|
| | | Meas. | RT | RT (ave) | 3GPP | $\|\Delta_1\|$ | $\|\Delta_2\|$ | $\|\Delta_3\|$ | Meas. | RT | RT (ave) | 3GPP | $\|\Delta_1\|$ | $\|\Delta_2\|$ | $\|\Delta_3\|$ |
| 6.75 | LOS | 1.79 | 1.86 | 1.86 | 2.19 | 0.07 | 0.07 | 0.40 | 2.57 | 1.01 | 1.06 | 4.00 | 1.56 | 1.51 | 1.43 |
| | NLOS | 2.56 | 2.66 | 2.82 | 3.19 | 0.10 | 0.26 | 0.63 | 6.53 | 11.73 | 16.30 | 8.20 | 5.20 | 9.77 | 1.67 |
| 16.95 | LOS | 1.85 | 1.88 | 1.88 | 2.10 | 0.03 | 0.03 | 0.25 | 4.05 | 1.11 | 1.14 | 4.00 | 2.94 | 2.91 | 0.05 |
| | NLOS | 2.59 | 2.73 | 2.85 | 3.19 | 0.14 | 0.26 | 0.60 | 8.78 | 12.01 | 14.36 | 8.20 | 3.23 | 5.58 | 0.58 |

$|\Delta_1|$ = |Meas. − RT|; $|\Delta_2|$ = |Meas. −RT(ave)|; $|\Delta_3|$ = |Meas. −3GPP|. "RT(ave)" denotes statistics computed by averaging the simulated PDP over 25 RX positions: the calibrated point plus offsets of 1, 2, and 3 λ in eight horizontal directions (±x, ±y, and the four diagonals). Data sources include Measurements (Meas.)[39], NYURay RT simulations (RT), spatially averaged NYURay RT simulations (RT(ave)), and the 3GPP UMi CI model[61]. Statistics are derived from 18 non-outage T-R locations (7 LOS, 11 NLOS) with T-R separations ranging from 40 m to 880 m.

filtering achieves substantial distributional improvements with KS statistic reductions of 13.5% at both frequencies, transitioning from statistically significant differences (KS = 0.4444, $p$ = 0.0560, indicating distributions likely differ) to statistically indistinguishable distributions (KS = 0.3846, $p$ = 0.2999, indicating distributions statistically equivalent) through systematic removal of 5 outlier pairs of T-R locations. The filtered CDFs exhibit substantially improved alignment between RT simulations and measurements, confirming that deterministic modeling accurately captures temporal dispersion characteristics when outlier pairs are systematically removed. Accurate DS prediction enables precise intersymbol interference estimation critical for adaptive equalization design and optimal symbol timing recovery in wideband systems[13]. Outlier identification is performed using a systematic filtering methodology, as detailed in Fig. 7 and described in Section IV.

**Comparison of measured and RT simulated angular spread:** Angular spread validation quantified spatial dispersion characteristics by extracting azimuth and zenith spreads from omnidirectional power angular spectra (PAS) synthesized from directional measurement sweeps. The methodology computed RMS angular spreads for arrival (ASA, ZSA) and departure (ASD, ZSD) directions, providing essential spatial channel characteristics for MIMO system design.

Statistical analysis compared measured angular spreads with NYURay predictions using absolute differences and distributional similarity metrics. KS tests assessed whether RT-predicted and measured distributions originate from the same underlying statistical process.

Table 7 presents comprehensive validation results revealing systematic patterns in angular spread prediction. The analysis demonstrates consistent underestimation across all angular spread parameters, with particularly notable variations in elevation spreads reflecting methodological differences between measurement and simulation approaches.

Application of the combined outlier filtering methodology significantly improves distributional agreement between RT simulations and measurements across all analyzed channel parameters. The filtering approach successfully identifies and removes RT-measurement pairs exhibiting extreme discrepancies, resulting in substantially improved statistical similarity as quantified through KS test analysis[75]. The KS statistic measures the maximum difference between cumulative distribution functions, where smaller values indicate better distributional matching essential for accurate system design predictions.

Figure 5 demonstrates the enhanced statistical validation for omnidirectional RMS ASA through outlier filtering. The RMS ASA validation demonstrates substantial improvement, with KS statistics reduced by 28.0% at 6.75 GHz (from 0.278 to 0.200) and increased by 49.7% at 16.95 GHz (from 0.167 to 0.250), while p-values improve from 0.503 to 0.994 at 6.75 GHz and from 0.972 to 0.980 at 16.95 GHz, transitioning from moderate statistical differences to excellent distributional agreement following removal of 7 outlier pairs. The filtered CDFs show excellent alignment between RT simulations and measurements, with median values achieving close agreement (RT: 14.4°, measured: 14.6° at 6.75 GHz; RT: 13.7°, measured: 10.5° at 16.95 GHz). Accurate RMS ASA prediction directly impacts MIMO beamforming performance, antenna array design optimization, and spatial multiplexing capacity estimation for massive MIMO deployments[76].

Figure 6 presents the enhanced validation results for omnidirectional RMS ASD following systematic outlier filtering. The filtering methodology produces frequency-dependent improvements, with KS statistics reduced by 11.9% at 6.75 GHz and 7.5% at 16.95 GHz, enhancing distributional similarity crucial for transmit beamforming optimization and interference coordination algorithms. The enhanced agreement demonstrates that deterministic RT modeling successfully captures the statistical characteristics of spatial dispersion when appropriate data processing methodologies are applied. Accurate ASD characterization enables optimal precoding matrix design and transmission strategy adaptation for multi-user MIMO systems[77].

RMS Zenith angular spread validation reveals the impact of both measurement methodology and environmental modeling on elevation

**Table 6 | RMS DS characteristics ($\mathbb{E}(\cdot)$ in ns) at 6.75 GHz and 16.95 GHz for the UMi environment in Downtown Brooklyn, NY around the NYU Brooklyn campus[39], with comparison to RT, spatially averaged RT (RT(ave))**

| Condition | 6.75 GHz | | | | | | | 16.95 GHz | | | | | | |
|---|---|---|---|---|---|---|---|---|---|---|---|---|---|---|
| | Meas. | RT | RT (ave) | 3GPP | Delta | | | Meas. | RT | RT (ave) | 3GPP | Delta | | |
| | | | | | $\|\Delta_1\|$ | $\|\Delta_2\|$ | $\|\Delta_3\|$ | | | | | $\|\Delta_1\|$ | $\|\Delta_2\|$ | $\|\Delta_3\|$ |
| $\mathbb{E}$( LOS ) (ns) | 62.8 | 24.7 | 24.5 | 52.7 | 38.1 | 38.3 | 10.1 | 46.5 | 21.6 | 21.3 | 42.9 | 24.9 | 25.2 | 3.6 |
| $\mathbb{E}$( NLOS ) (ns) | 75.6 | 52.9 | 45.1 | 111.1 | 22.7 | 30.5 | 35.5 | 65.8 | 53.6 | 50.9 | 96.7 | 12.2 | 14.9 | 30.9 |

$|\Delta_1|$ = |Meas. − RT|; $|\Delta_2|$ = |Meas. − RT(ave)|; $|\Delta_3|$ = |Meas. − 3GPP|.

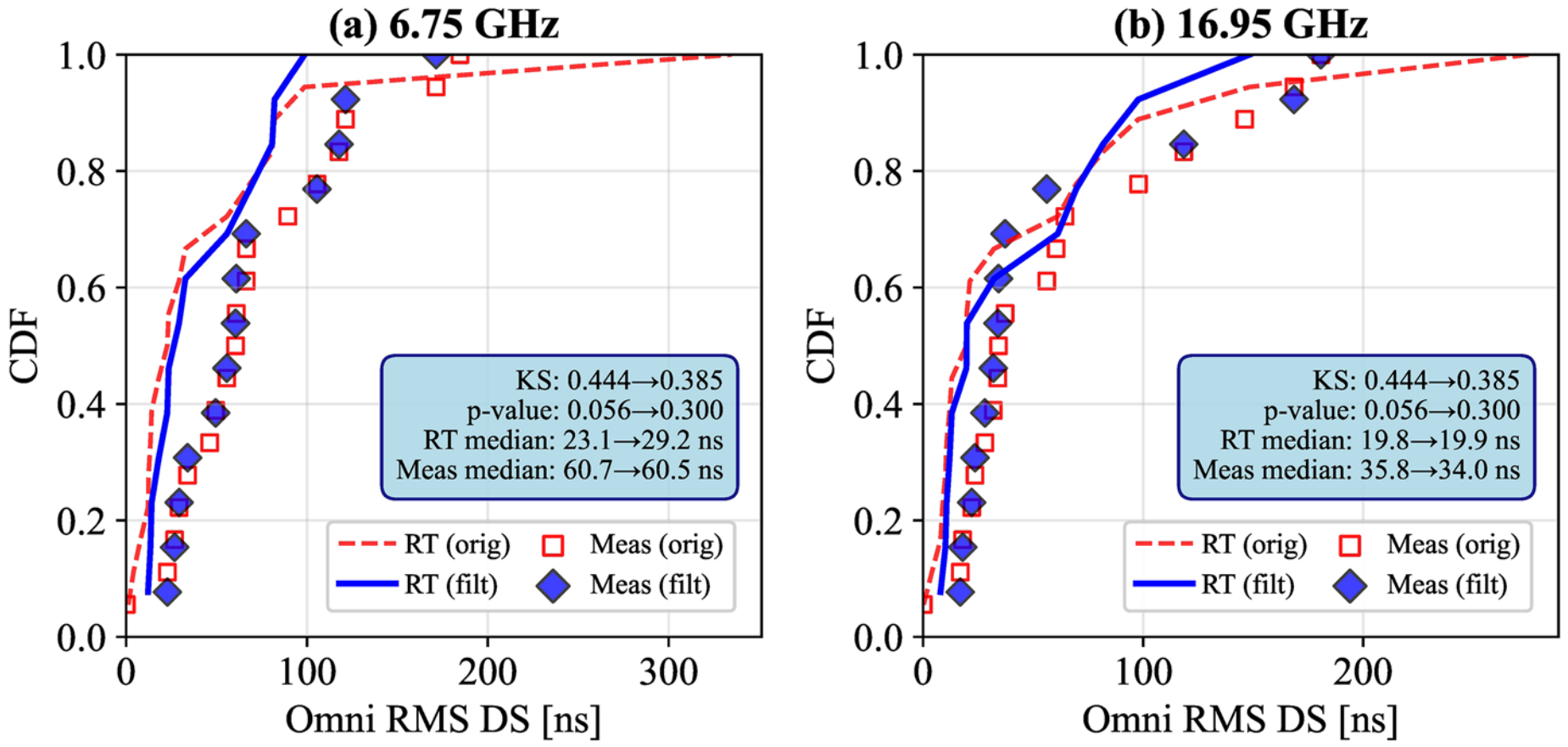

**Fig. 4 | Statistical validation of omnidirectional RMS delay spread across 18 T-R locations in the NYU Brooklyn campus UMi environment.** (**a**) 6.75 GHz and (**b**) 16.95 GHz. RT data shown as lines (red dashed line for original DS data, blue solid line for filtered DS data), measured data as markers (red empty squares for original, blue filled diamonds for filtered). Measured data points are from ref. 39.

dispersion prediction. As shown in Table 7, RT simulations systematically underestimate both ZSA and ZSD compared to measurements, with the largest discrepancies observed in LOS scenarios. For example, at 6.75 GHz, measured LOS ZSA is 16.6° while RT predicts 2.9°, and measured LOS ZSD is 10.6° versus RT's 2.8°. Similar trends are observed at 16.95 GHz and in NLOS conditions, though the absolute spreads are smaller. The underestimation of ZAS and ZSD is attributed to limited elevation angle measurements during the field campaign. Better agreement at 16.95 GHz than at 6.75 GHz suggests fewer elevation measurements are needed at higher frequencies to synthesize accurate omnidirectional characteristics, consistent with Sun et al. [60], who showed that millimeter-wave and terahertz frequencies require sparser elevation sampling than lower frequencies.

The underestimation of RMS ZSA and ZSD arises from several factors. First, the measurement system samples elevation angles using discrete antenna boresight steps, and interpolation is required to estimate the true zenith spread, potentially broadening the measured ZSA/ZSD. In contrast, RT simulations compute exact geometric angles for each multipath component, resulting in finer angular resolution but potentially missing diffuse or off-axis energy due to environmental simplifications. Second, the 3D environmental model may lack sub-wavelength features, rooftop details, and small scatterers that contribute to elevation dispersion in real environments. Third, the absence of diffuse scattering and dynamic elements (e.g., moving vehicles, foliage) in RT further limits elevation spread richness.

Despite these limitations, RT-predicted ZSA/ZSD values are generally consistent with 3GPP model assumptions for UMi scenarios, and the relative trends between LOS and NLOS are preserved. The systematic underestimation highlights the need for enhanced environmental modeling and improved measurement-simulation alignment in elevation, especially as elevation domain MIMO and 3D beamforming become more prevalent in next-generation wireless systems[76].

The enhanced statistical validation demonstrates that environmental modeling simplifications primarily affect unfiltered comparisons rather than fundamental RT accuracy. When systematic outlier filtering removes problematic measurement-simulation pairs, the remaining data achieves excellent statistical agreement with p-values exceeding 0.05 (conventional threshold for statistical equivalence), confirming that deterministic RT modeling accurately captures the statistical characteristics of upper mid-band propagation when appropriate data processing methodologies are applied.

## Channel validation summary and performance assessment

The comprehensive validation analysis demonstrates both the capabilities and limitations of deterministic RT modeling for upper mid-band propagation prediction. NYURay achieves exceptional path loss accuracy with PLE differences between 0.03 and 0.14 from measurements after location calibration, confirming the capability to capture dominant large-scale propagation mechanisms essential for coverage prediction and network planning applications. The strong quantitative agreement in first-order channel statistics validates NYURay for practical wireless system design tasks including base station placement optimization and interference analysis.

However, validation reveals systematic limitations in multipath richness characterization, with RT simulations consistently underestimating DS and AS compared to measurements across both LOS and NLOS conditions. The DS underestimation averages 38.1 ns (61%) for LOS and 22.7 ns (30%) for NLOS scenarios, while AS underestimation ranges from 7 to 15° across all angular parameters. The multipath deficiency primarily stems from environmental model simplifications, including absence of sub-wavelength scatterers, limited architectural detail representation, and omission of dynamic environmental factors such as foliage movement, pedestrian activity, and moving vehicles that contribute to measured multipath richness.

The observed discrepancies directly impact second-order channel statistics critical for advanced wireless system design. Underestimated DS affects adaptive equalization design and channel estimation algorithms, while reduced AS predictions impact MIMO spatial correlation modeling, beamforming performance prediction[78], and antenna array optimization. Accurate temporal and spatial dispersion characterization becomes

**Table 7 | Comparison of expected RMS AS characteristics ($\mathbb{E}(\cdot)$ in degrees) for the UMi environment in Downtown Brooklyn, NY around the NYU Brooklyn campus**

| Metric (°) | Condition | 6.75 GHz | | | | | | | 16.95 GHz | | | | | | |
|---|---|---|---|---|---|---|---|---|---|---|---|---|---|---|---|
| | | Meas. | RT | RT (ave) | 3GPP | Delta (°) | | | Meas. | RT | RT (ave) | 3GPP | Delta (°) | | |
| | | | | | | $\|\Delta_1\|$ | $\|\Delta_2\|$ | $\|\Delta_3\|$ | | | | | $\|\Delta_1\|$ | $\|\Delta_2\|$ | $\|\Delta_3\|$ |
| Omni RMS ASA | LOS | 16.8 | 7.9 | 8.6 | 50.4 | 8.9 | 8.2 | 33.6 | 18.2 | 7.2 | 7.9 | 47.3 | 11.0 | 10.3 | 29.1 |
| | NLOS | 25.6 | 13.4 | 9.6 | 62.8 | 12.2 | 16.0 | 37.2 | 19.1 | 12.8 | 8.7 | 59.5 | 6.3 | 10.4 | 40.4 |
| Omni RMS ASD | LOS | 32.1 | 16.8 | 16.8 | 17.5 | 15.3 | 15.3 | 14.6 | 23.7 | 14.7 | 14.7 | 17.1 | 9.0 | 9.0 | 6.6 |
| | NLOS | 37.6 | 17.4 | 12.7 | 25.9 | 20.2 | 24.9 | 11.7 | 31.4 | 17.7 | 14.7 | 22.4 | 13.7 | 16.7 | 9.0 |
| Omni RMS ZSA | LOS | 16.6 | 2.9 | 2.8 | 4.8 | 13.7 | 13.8 | 11.8 | 10.0 | 2.8 | 2.8 | 4.4 | 7.2 | 7.2 | 5.6 |
| | NLOS | 12.6 | 1.4 | 1.3 | 8.7 | 11.2 | 11.3 | 3.9 | 8.5 | 1.3 | 1.2 | 8.3 | 7.2 | 7.3 | 0.2 |
| Omni RMS ZSD | LOS | 10.6 | 2.8 | 2.8 | 2.1 | 7.8 | 7.8 | 8.5 | 7.0 | 2.4 | 2.4 | 2.1 | 4.6 | 4.6 | 4.9 |
| | NLOS | 8.8 | 1.5 | 1.4 | 1.5 | 7.3 | 7.4 | 7.3 | 7.5 | 1.4 | 1.2 | 1.5 | 6.1 | 6.3 | 6.0 |

Values are derived from NYU WIRELESS Measurements (Meas.)[40,52], NYURay RT Simulations (RT), Spatially Averaged RT Simulations (RT (ave)), and standard 3GPP Models[61] at 6.75 GHz and 16.95 GHz under LOS/NLOS conditions using 18 non-outage locations.
$|\Delta_1| = |\text{Meas.} - \text{RT}|$; $|\Delta_2| = |\text{Meas.} - \text{RT(ave)}|$; $|\Delta_3| = |\text{Meas.} - \text{3GPP}|$. RT and RT(ave) statistics exclude zero AS values from simulation.

increasingly critical for massive MIMO deployments, millimeter-wave communications, and reconfigurable intelligent surface optimization, where precise channel statistics drive system performance[76,77].

The limitations highlight fundamental challenges in balancing computational efficiency with environmental modeling fidelity. While RT accurately captures geometric propagation paths through major reflections and diffractions, comprehensive multipath characterization requires enhanced environmental models incorporating detailed surface roughness, sub-wavelength objects, and dynamic scattering mechanisms. Future research directions include developing automated high-fidelity environment generation from LiDAR and photogrammetry data, and machine learning techniques for intelligent environmental model enhancement. Advanced material characterization databases and computational optimization will enable more accurate multipath prediction while maintaining practical simulation times for real-world deployment scenarios.

The observed DS/AS underestimation may be attributed to the lack of dynamic environmental factors in static RT simulations. Static-environment RT inherently misses extra multipath components and angular diversity created by temporally random vehicles and pedestrians, potentially yielding DS/AS values systematically lower than measured values. Literature evidence supports this hypothesis, as Yan et al.[79] demonstrate DS approximately triples from 6–7 ns to 22–23 ns as vehicular traffic increases from low to full flow at 22–23 GHz, clearly illustrating the significant impact of dynamic scatterers on temporal dispersion characteristics. Highway RMS-DS scales strongly with traffic density, while horizontal angular spread (ASA/ASD) in urban environments grows with base station height and vehicle presence.

For pedestrian movement effects, Gapeyenko et al.[80] show mobile-blocker processes using Poisson arrivals through precomputed blockage zones, where non-blocked intervals follow exponential distributions and blocked durations model busy periods. Such approaches could enable realistic short-time DS/AS excursions without explicitly simulating every pedestrian trajectory, potentially addressing some of the observed underestimation. These validation findings and their implications for future wireless system design are further explored in the Discussion section.

For vehicular environments, the literature suggests that moving vehicles significantly affect DS and horizontal AS characteristics. Yan et al.[79] demonstrate that moving vehicles dominate DS and horizontal AS on highways, with multi-beam switching producing 10–20 dB power differences across beams during overtaking maneuvers, which could contribute to the angular spread underestimation observed in static simulations.

Human presence modeling research indicates effects beyond classic blockage scenarios. MacCartney et al.[81] demonstrate through four-state piecewise linear modeling that signal strength decay and rise times are asymmetric for blockage events, with mean signal attenuations inversely proportional to antenna HPBW. Their measurements show 7° and 60° HPBW antennas resulted in mean signal fades of 15.8 dB and 11.5 dB, respectively, highlighting the importance of antenna pattern considerations in blockage modeling. Building on these blockage characterizations, Bhardwaj et al.[82] demonstrate that accounting for human-body reflections before and after crossing events, combined with diffraction during blockage, achieves 5–20 dB improved fading prediction compared to classic blockage-only models. Such approaches capture oscillatory pre-shadow energy effects and time-dispersion characteristics that static models cannot represent.

Rodriguez-Corbo et al.[83] present hybrid deterministic-stochastic frameworks that maintain computational efficiency while capturing dynamic environmental effects. Their approaches involve temporal binning over short windows and hybrid acceleration techniques, demonstrating potential pathways for addressing the limitations observed in static RT modeling.

Future NYURay enhancements will consider integrating dynamic environmental modeling approaches to address the DS/AS underestimation and achieve more comprehensive channel characterization that captures the temporal and spatial dispersion effects essential for next-generation wireless system design.

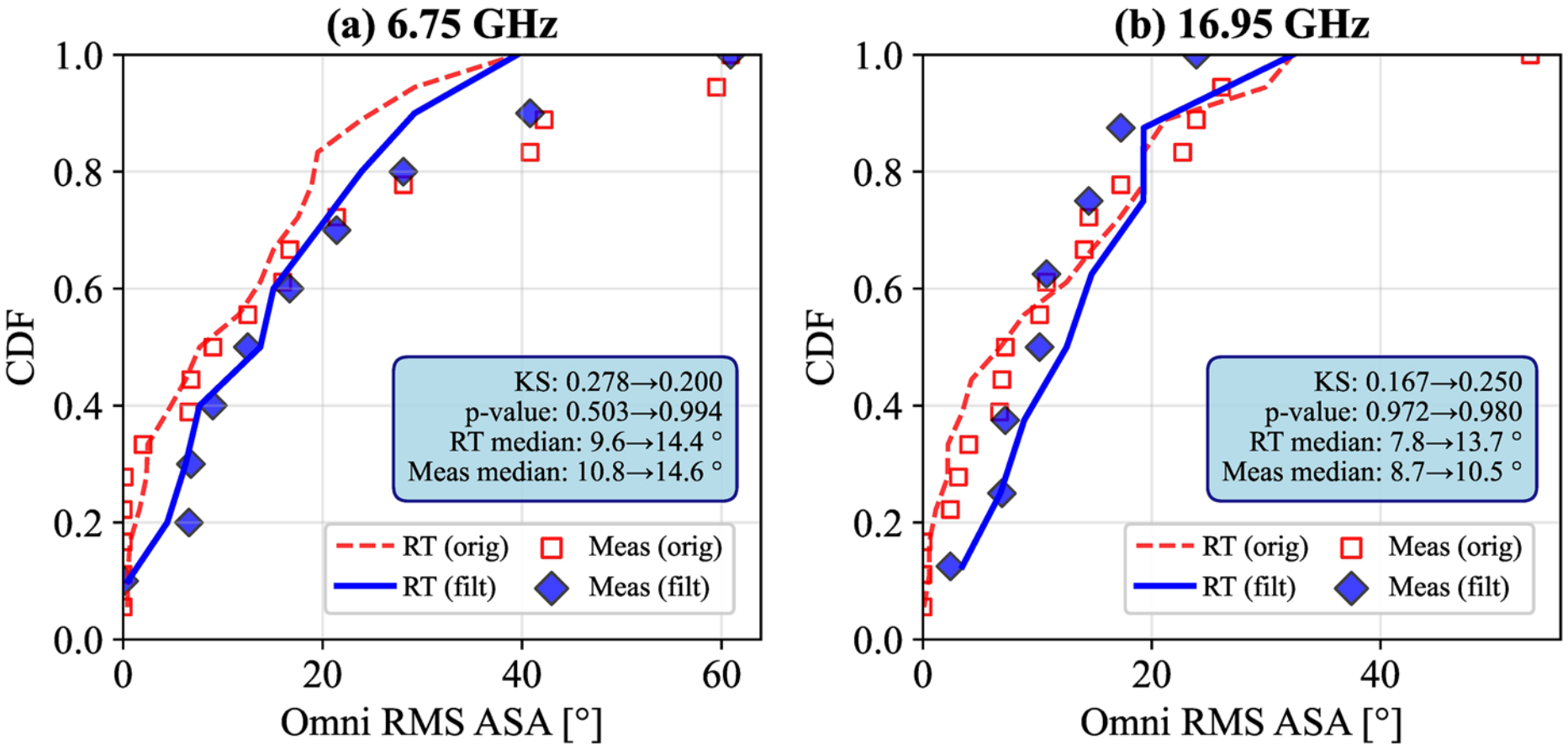

**Fig. 5 | Statistical validation of omnidirectional RMS azimuth spread of arrival across 18 T-R locations in the NYU Brooklyn campus UMi environment.** (**a**) 6.75 GHz and (**b**) 16.95 GHz. RT data shown as lines (red dashed line for original ASA data, blue solid line for filtered ASA data), measured data as markers (red empty squares for original, blue filled diamonds for filtered). Measured data points are from ref. 39.

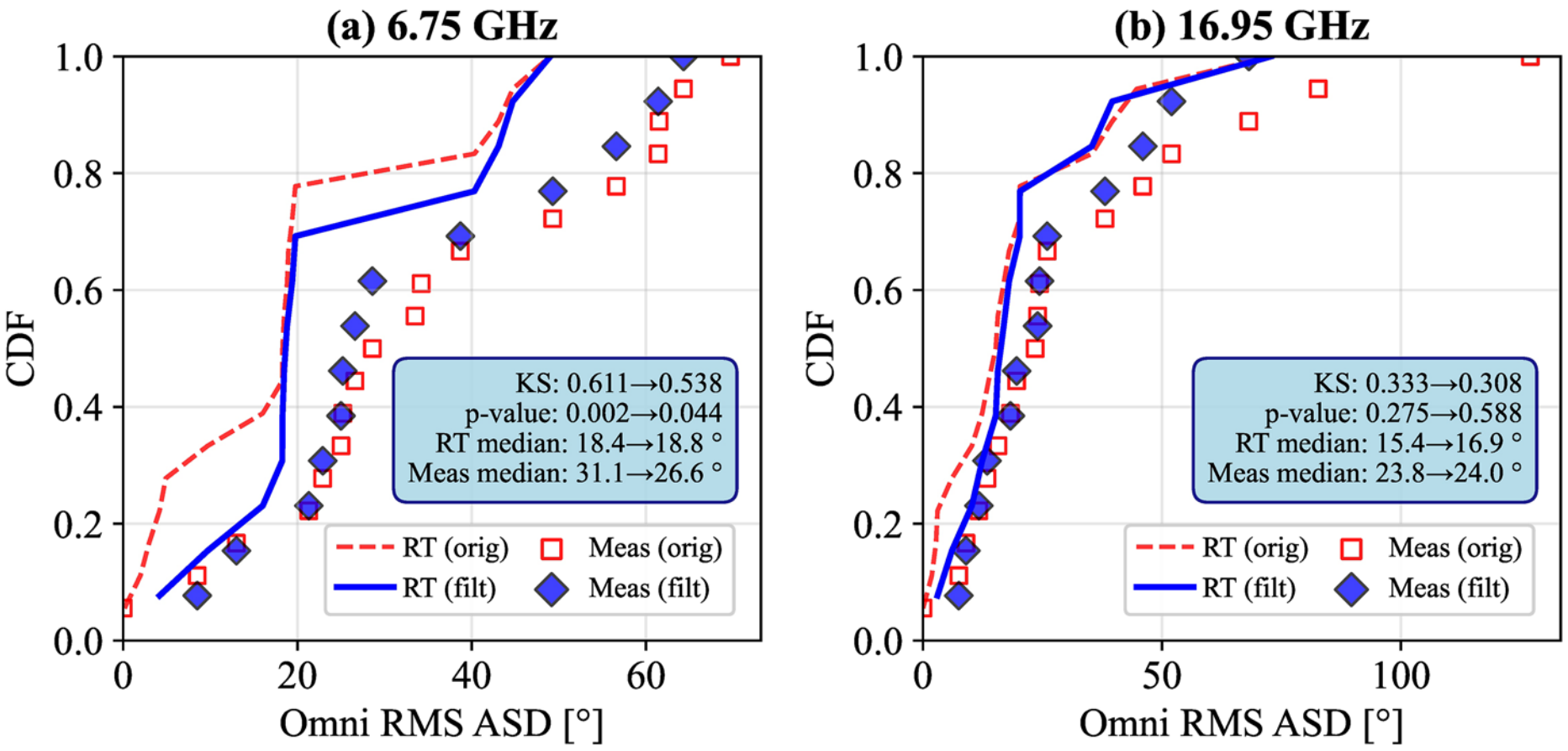

**Fig. 6 | Statistical validation of omnidirectional RMS azimuth spread of departure across 18 T-R locations in the NYU Brooklyn campus UMi environment.** (**a**) 6.75 GHz and (**b**) 16.95 GHz. RT data shown as lines (red dashed line for original ASD data, blue solid line for filtered ASD data), measured data as markers (red empty squares for original, blue filled diamonds for filtered). Measured data points are from ref. 39.

## Discussion

This paper validates NYURay by comparing RT-simulated path loss, delay spread, and angular spread against wideband (1 GHz bandwidth) directional horn antenna measurements at 6.75 GHz and 16.95 GHz, where measured directional PDPs were synthesized into omnidirectional PDPs[60] for direct comparison with isotropic RT predictions across 18 T-R locations spanning 40 m to 880 m in Downtown Brooklyn, NY. This comprehensive validation establishes a foundational capability for deterministic wireless channel prediction in upper mid-band frequencies, addressing critical needs for 6G network deployment and wireless digital twin development. While achieving exceptional path loss accuracy with PLE deviations of 0.03-0.14 from measurements, the systematic underestimation of multipath richness (DS underestimation of 61% for LOS and 30% for NLOS scenarios; AS underestimation of 7–15 degrees) reveals fundamental trade-offs between computational efficiency and environmental modeling fidelity. The location calibration methodology successfully mitigates GPS uncertainties, demonstrating a 42.28% improvement in location accuracy for LOS scenarios and a 13.52% improvement for NLOS scenarios, establishing the critical foundation for accurate site-specific channel validation between simulated and measured results.

NYURay predictions are validated against the 3GPP TR 38.901 channel model[61], and provide insights to 6G standardization effort at upper mid-band from 7 to 24 GHz[15], which call for measurement-validated, site-specific channel models capable of producing reliable large-scale propagation parameters, demonstrates strong agreement between ray-tracing simulations and standardized large-scale channel characteristics. Detailed comparison shows path loss exponents simulated using NYURay align within 5% of the 3GPP UMi CI model. Our measured path loss exponents (LOS 1.79 at 6.75 GHz, 1.85 at 16.95 GHz; NLOS 2.56 at 6.75 GHz, 2.59 at 16.95 GHz) also align within 5% of the 3GPP model in[61]. Overall, the close agreement between simulation, measurement, and the 3GPP baseline supports the accuracy of NYURay for site-specific upper mid-band prediction in UMi environments. NYURay's support for enhanced site-specific modeling requirements enables deterministic channel prediction essential for dense small cell deployments, massive MIMO optimization, and dynamic spectrum management. Real-time prediction capabilities for 6G digital twins meet the computational efficiency envisioned for next-generation wireless systems. The demonstrated calibration methodology contributes to upper mid-band propagation modeling standardization in 3GPP Release 19 and beyond.

By accurately predicting real-world measured propagation, NYURay enables impactful use across emerging 6G applications. Accurate path loss prediction directly translates to optimized coverage planning, significantly reducing infrastructure costs through precise base station placement while eliminating coverage gaps. Early interactive systems demonstrated the value of computational tools for coverage optimization in complex indoor environments[6]. Enhanced angular spread characterization enables massive MIMO systems to achieve improved spatial multiplexing performance, with more accurate beamforming substantially reducing interference in dense deployments[76,78,84]. For industrial IoT, precise delay spread prediction ensures robust ultra-reliable low-latency communication design, enabling the stringent latency requirements critical for autonomous manufacturing systems. Smart city networks leverage deterministic channel knowledge for dynamic spectrum allocation and interference coordination, achieving improved spectral efficiency compared to statistical models. Validated RT simulators enable high-accuracy synthetic dataset generation for indoor navigation and localization applications[85], but reliable RT calibration and ML training still require high-quality wideband field measurements as ground truth. Once calibrated, synthetic data can substantially reduce the need for repeated large-scale measurement campaigns. The standardized point-data format facilitates seamless integration with AI-driven network optimization tools, enabling autonomous infrastructure management where real-time channel predictions drive adaptive resource allocation and self-healing network operations. Furthermore, the structured point-data format enables efficient data sharing across research communities and provides standardized datasets optimized for machine learning training, accelerating collaborative development of AI-enhanced wireless communication systems.

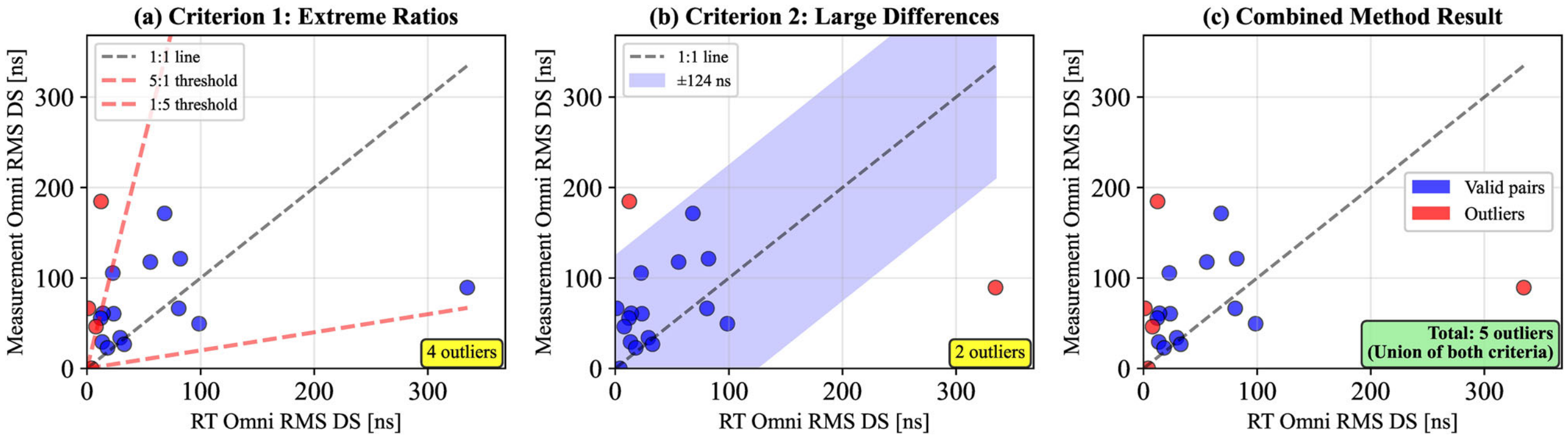

**Fig. 7 | Combined filtering method demonstration using delay spread at 6.75 GHz. a** shows extreme ratio criterion identifying outliers when ratios fall below 0.2 or exceed 5.0. **b** shows absolute difference criterion flagging outliers above the 90th percentile. **c** shows combined approach identifying all problematic pairs. Red points indicate outliers for removal, blue points show retained valid data.

Future research directions prioritize AI-enhanced 3D environmental modeling through deep learning techniques that automatically generate high-fidelity digital environments from multimodal sensor data, including LiDAR, photogrammetry, and street-view imagery. Real-time RT simulation capabilities will leverage neuromorphic computing and advanced GPU architectures to achieve sub-second channel prediction updates, enabling true wireless digital twins with continuous environment-channel co-evolution tracking. Hybrid channel modeling approaches will integrate physics-based RT with generative AI models to synthesize missing multipath components, combining deterministic accuracy with stochastic richness. The convergence of advanced modeling technologies toward intelligent, self-updating digital twins represents the pathway to autonomous 6G networks capable of predictive optimization and zero-touch operation in dynamic environments. The technical methods enabling these validation results and analyses are detailed in the following section.

## Methods

### Outlier detection and CDF evaluation methods

RT simulation accuracy requires systematic outlier identification and removal to improve agreement between simulations and measurements. The outlier detection methodology identifies and removes problematic RT-measurement pairs through statistical filtering approaches. Let $\mathbf{x}_{RT} = \{x_{RT}^{(i)}\}_{i=1}^{N}$ and $\mathbf{x}_{meas} = \{x_{meas}^{(i)}\}_{i=1}^{N}$ denote paired RT and measurement values for channel parameters, including DS, ASA, and ASD, where $N$ represents the total number of T-R pairs.

The ratio-based analysis defines the measurement-to-RT ratio for each paired observation

$$r_i = \frac{x_{meas}^{(i)}}{x_{RT}^{(i)} + \epsilon} \tag{21}$$

where $\epsilon = 10^{-6}$ prevents division by zero. The ratio quantifies relative differences between measurement and simulation values.

The combined filtering method employs multiple criteria to identify outliers based on absolute and relative differences, as illustrated in Fig. 7. The filtering process follows five systematic steps. First, the measurement-to-RT ratio $r_i = x_{meas}^{(i)}/x_{RT}^{(i)}$ is calculated for each paired observation. Second, Criterion 1 examines extreme ratios where $r_i < 0.2$ indicates RT values 5× larger than measurements, and $r_i > 5.0$ indicates measurement values 5× larger than RT predictions, as shown in Fig. 7a. Third, Criterion 2 evaluates absolute differences

$$d_i = \left| x_{meas}^{(i)} - x_{RT}^{(i)} \right| \tag{22}$$

where pairs with differences exceeding the 90th percentile are flagged (Fig. 7b)

$$\mathcal{C}_1 = \{i : d_i > P_{90}(\mathbf{d})\} \tag{23}$$

The extreme ratio criterion identifies pairs

$$\mathcal{C}_2 = \{i : r_i < 0.2 \text{ or } \mathrm{r_i} > 5.0\} \tag{24}$$

The ratio thresholds $r_i < 0.2$ and $r_i > 5.0$ are based on physical reasoning and empirical validation. Values outside the 0.2–5.0 range indicate one prediction is at least 5 times larger than the other, corresponding to approximately 7 dB difference for power-related quantities. Such large discrepancies significantly exceed typical combined measurement and simulation uncertainties (1–3 dB for path loss, 2–4 dB for temporal parameters) and likely indicate fundamental issues including missing environmental features in the RT model, incorrect material properties, or unmodeled propagation mechanisms.

Fourth, if either criterion is satisfied, the pair is marked as an outlier. The final outlier set combines both criteria

$$\mathcal{O}_{combined} = \mathcal{C}_1 \cup \mathcal{C}_2 \tag{25}$$

as demonstrated in Fig. 7c.

Fifth, outliers are removed from the dataset, and filtered CDFs undergo comparison to assess improvement. The combined method captures different error types through physically meaningful thresholds, adapts to data characteristics through percentile-based criteria, and removes only problematic pairs while preserving valid measurements.

Zero values indicating measurement outages are excluded before analysis

$$\mathcal{I}_{valid} = \left\{ i : x_{RT}^{(i)} > 0 \text{ and } x_{meas}^{(i)} > 0 \right\} \tag{26}$$

Filtering effectiveness is evaluated using the KS test statistic[75]

$$D_{KS} = \sup_{x} |F_{RT}(x) - F_{meas}(x)| \quad (27)$$

where $F_{RT}(x)$ and $F_{meas}(x)$ represent empirical cumulative distribution functions of filtered RT and measurement data. Statistical significance assessment employs the two-sample KS test with null hypothesis $H_0$ where $F_{RT} = F_{meas}$ at significance level $\alpha = 0.05$. Reduced KS statistic $D_{KS}$ and increased p-value after filtering indicate improved agreement between RT simulations and measurements.

## Data availability

Raw channel measurement data at 6.75 GHz and 16.95 GHz for the urban microcell environment can be obtained through the NYU WIRELESS industry affiliate program. The point-data tables for UMi campaign are included within[39].

## Code availability

The code for this study is not publicly available but might be made available, conditioned upon approval of the NYU Technology Opportunities and Ventures (TOV) office.

## Acknowledgements

This work was supported by the NYU WIRELESS Industrial Affiliates Program for this research. This work was also partially supported by the PhD Fellowship from NYU Tandon School of Engineering and by the National Science Foundation under Grant No. 2234123.

## Author contributions

M.Y. developed the simulator, implemented algorithms, conducted simulations and analysis, and wrote the manuscript. D.S. designed measurements and developed data processing. P.M. contributed to 3D modeling. G.Q. assisted with data analysis. T.R. supervised the research and reviewed the manuscript. All authors read and edited the paper.

## Competing interests

The authors declare no competing interests.

## Additional information

**Correspondence** and requests for materials should be addressed to Mingjun Ying or Theodore S. Rappaport.